\RequirePackage{ifpdf}
\documentclass[hyper]{JHEP3}

\pdfoutput=1
\usepackage{epsfig}
\usepackage{cite}
\usepackage{graphicx}
\usepackage{subfigure}
\usepackage{inputenc}
\usepackage{amsmath}
\usepackage{diagbox}

\newcommand{\beq}{\begin{equation}}
\newcommand{\eeq}{\end{equation}}
\newcommand{\beqn}{\begin{eqnarray}}
\newcommand{\eeqn}{\end{eqnarray}}
\def\df{{\rm d}}

\newcommand{\eq}[1]{Eq.~\eqref{eq:#1}}
\newcommand{\eqs}[2]{Eqs.~\eqref{eq:#1} and \eqref{eq:#2}}
\newcommand{\vev}[1]{\langle{#1}\rangle}
\def\a{\alpha}

\def\nn{\nonumber\\}
\def\pd{\partial}
\def\SU{{\rm SU}} 
\def\U{{\rm U}} 

\title{Standard Model Fragmentation Functions at Very High Energies}

\author{Christian W.~Bauer$^{a}$, Davide Provasoli$^a$ and Bryan R.~Webber$^b$\\
   $^a$Ernest Orlando Lawrence Berkeley National Laboratory, University of California, Berkeley, CA 94720, USA\\
    $^b$University of Cambridge, Cavendish Laboratory, J.J.\ Thomson Avenue, Cambridge, UK\\
        E-mail: \email {cwbauer@lbl.gov}, \email{davideprovasoli@berkeley.edu}, \email{webber@hep.phy.cam.ac.uk}
        }

\received{\today}               
\accepted{\today}               

\preprint{Cavendish-HEP-18/12}

\abstract{We compute the leading-order evolution of parton
  fragmentation functions for all the Standard Model fermions and
  bosons up to energies far above the electroweak scale, where
  electroweak symmetry is restored. We discuss the difference between
  double-logarithmic and leading-logarithmic resummation, and show
  how the latter can be implemented through a scale choice in the
  SU(2) coupling.  We present results for a wide range of
  partonic center-of-mass energies, including the polarization of
  fermion and vector boson fragmentation functions
  induced by electroweak evolution.  
}
\keywords{Standard Model, Parton Distributions}

\begin{document} 

\section{Introduction}
\label{sec:intro}
It is a well known fact that electroweak corrections to hard processes at proton or electron colliders contain logarithmically enhanced contributions of the form $\alpha^n L^{2n}$, 
where $L = \ln (q/ m_V)$, $q$ being the hard process scale and $m_V\sim m_{W/Z}$. This is the case even for observables that are completely inclusive over the final state, and can be traced back 
to the fact that the initial state protons are not singlets under the SU(2) gauge group. Due to this double-logarithmic scaling, the convergence of electroweak perturbation 
theory becomes worse as the center-of-mass energy increases, and
ultimately breaks down completely, namely when $\alpha L^2 \sim 1$. To obtain 
reliable predictions at these energy scales requires a reorganization of the perturbative expansion such that these large logarithms are resummed to all orders
in perturbation theory.

Most of the studies of electroweak logarithms have considered completely exclusive observables, such that the final state is fixed. In this case the 
only contributions to logarithmically enhanced electroweak corrections arise from the virtual exchange of massive gauge bosons. The mass of the gauge boson 
regulates the IR divergences present in massless gauge theories, giving rise to the logarithmic sensitivity on $m_V$. These electroweak Sudakov logarithms 
have been studied for a long time~\cite{Kuhn:1999nn,Fadin:1999bq,Beccaria:2000jz,Hori:2000tm,Denner:2000jv,Denner:2001gw,Melles:2001ye,Beenakker:2001kf,
Denner:2003wi,Pozzorini:2004rm,Feucht:2004rp,Jantzen:2005xi,Jantzen:2005az,Jantzen:2006jv,Chiu:2007yn,Chiu:2008vv,Manohar:2012rs}, and a systematic 
way to resum them using soft-collinear effective theory (SCET)\cite{Bauer:2000ew,Bauer:2000yr,Bauer:2001ct,Bauer:2001yt} was developed in~\cite{Chiu:2007yn,Chiu:2008vv}.
Just as for massless theories, the real radiation of gauge bosons leads to infrared sensitivity, and therefore logarithmic sensitivity to $m_V$ is present in such real 
emission contributions as well.  An analogy with parton showers allowed the resummation of the enhanced corrections to leading logarithmic (LL) 
accuracy~\cite{Bauer:2016kkv}.

As already discussed, even fully inclusive observables contain double logarithmic sensitivity to the ratio $q / m_V$, due to the fact that the initial state is not an 
SU(2) singlet. For an observable that is completely inclusive over the final state, all logarithmically enhanced terms arise from initial-state radiation of $W$ 
bosons. To LL accuracy, the large logarithms arise from emissions of heavy gauge bosons that are both collinear and soft, and are described by the DGLAP evolution 
of parton distribution functions~\cite{Ciafaloni:1998xg,Ciafaloni:1999ub,Ciafaloni:2000df,Ciafaloni:2000gm,Ciafaloni:2000rp,Ciafaloni:2001vu,Ciafaloni:2001mu,
Ciafaloni:2003xf,Ciafaloni:2005fm,Ciafaloni:2006qu,Ciafaloni:2008cr,Ciafaloni:2009mm,Ciafaloni:2010ti,Forte:2015cia,Mangano:2016jyj}, where one needs the 
full set of particles in the Standard Model. These DGLAP equations were first derived in~\cite{Ciafaloni:2005fm}, and the phenomenology of this DGLAP evolution 
in the complete SM was studied
in~\cite{Bauer:2017isx,Bauer:2017bnh}. As will be shown in this paper, while the DGLAP evolution presented in~\cite{Bauer:2017isx,Bauer:2017bnh} was only accurate to double-logarithmic level, the full LL structure can be obtained for such completely inclusive observables through an appropriate scale choice in the SU(2) coupling constant. 

Most realistic observables, however, contain a final state which is neither fully inclusive nor fully exclusive. The results
of~\cite{Manohar:2018kfx} allow one to obtain
NLL predictions where a subset of the final state particles is fixed,
while being inclusive over the emission of additional particles. So for example, they allow one to compute
the cross section of the process $p p \to e^+ e^- + X$, where $X$ denotes additional particles in the final state. For the most general case, where one wants to 
include additional final state particles only partially (for example only in a given kinematic range, or only those that decay in a particular way) one needs to use an 
electroweak parton shower, which generates an arbitrary final state. If formulated correctly, such a parton shower will resum all LL electroweak Sudakov logarithms, 
and furthermore include many (but not all) of the NLL logarithms. A final-state parton shower including emissions from all interactions in the Standard Model 
was developed~\cite{Chen:2016wkt}, which also paid special attention to important threshold effects for longitudinal gauge bosons. 

To obtain the full NLL accuracy of~\cite{Manohar:2018kfx} requires
four types of input: The hard cross sections evaluated at the partonic
center-of-mass energy in the unbroken $\SU(3) \otimes \SU(2) \otimes
\U(1)$ Standard Model, the parton distributions functions (PDFs)
describing the collinear evolution of the initial-state particles,
the fragmentation functions (FFs) describing the collinear evolution
of the final-state particles, and a soft function describing the
wide-angle soft radiation. The collinear evolution needs to be
performed with the full gauge structure $\SU(3) \otimes \SU(2) \otimes
\U(1)$ and was discussed for the PDFs in detail
in~\cite{Bauer:2017isx,Bauer:2017bnh}. In this paper we will perform a
similar analysis for the FFs, including numerical results showing the
impact of the logarithmic terms. Our results can be used as one of the
inputs to~\cite{Manohar:2018kfx}, which allows full NLL
accuracy. However, at collider energies that are achievable with
current technologies one has the scaling $\alpha L\ll 1$, such that LL
accuracy, matched with fixed-order electroweak corrections as discussed
in~\cite{Bauer:2017bnh}, will be sufficient for most applications of interest. In this case one can omit the soft functions and use the hard cross sections only in combination with the collinear evolution of PDFs and FFs~\footnote{An analysis of the size of various contributions to the full NLL resummation in exclusive processes was performed in~\cite{Chiu:2009mg}.}

This paper is organized as follows: In Section~\ref{sec:fragLogs} we
discuss the form of the fragmentation function and their DGLAP
evolution with $q$. This discussion is correct to double logarithmic
accuracy, and we discuss in Section~\ref{sec:leading_log_2} how the
results can be modified to achieve full leading-logarithmic accuracy
through an appropriate scale choice of the SU(2) coupling
$\alpha_2$. After a brief discussion of some implementation details in
Section~\ref{sec:implementation}, we present the results for the
fragmentation functions in Section~\ref{sec:results}. Our conclusions
are given in Section~\ref{sec:Conclusions}, and in
Appendix~\ref{app:isospin} and~\ref{app:forward} we give details of an
isospin and CP basis that decouples parts of the DGLAP evolution and
the equations used in the forward evolution. 

\section{Resummation of collinear final-state logarithms}
\label{sec:fragLogs}
Electroweak logarithms arise from the exchange of massive gauge bosons in loops, or from the real radiation of massive gauge bosons. To LL accuracy, the only contributions are from gauge bosons that are collinear to one of the initial- or final-state particles. These are precisely the contributions that are contained in the DGLAP evolution of PDFs (for emissions collinear to initial-state particles) and FFs (for emissions collinear to final-state particles).

In the strong sector, the DGLAP equations only give rise to single logarithmic terms. This is because the 
limits where emissions are simultaneously soft and collinear cancel between virtual and real contributions 
to the DGLAP equations. This fact is easy to understand, since an arbitrarily soft emission of a gluon cannot
be observed experimentally, so the divergence associated with this emission needs to cancel against
the virtual contribution. This is different from the case of the soft emission of a $W$ boson, which 
can always be observed through the change of flavor (or SU(2) quantum
numbers) of the emitting particle. Thus, as long as a process is
sensitive to the SU(2) quantum numbers of the external particles, soft
radiation of $W$ bosons from these particles gives rise to double logarithms. 

Any observable at hadron or lepton colliders is sensitive to the SU(2) quantum numbers of the initial state, 
since the particles being collided are not SU(2) singlets. This leads to the important prediction that 
electroweak double logarithms are present for any observable, even if they are completely inclusive
over the final state. For observables where one identifies the SU(2) properties of the final state (for 
example by demanding to find two leptons of given flavors), additional double logarithms arise from
the collinear radiation off final state particles (even if one is completely inclusive over the momenta 
of said particles, and also over extra particles being radiated). These collinear logarithms can be resummed
by solving the DGLAP evolution of FFs, as we will now discuss.

DGLAP equations for FFs are very similar to those for PDFs, and the discussion of them
closely follows~\cite{Bauer:2017isx,Bauer:2017bnh}. We will therefore be relatively brief in this work, 
and refer the reader to the previous papers for more discussion.

Our solutions to the SM evolution equations are obtained in the approximation
of exact SU(3)$\times$SU(2)$\times$U(1) symmetry.  That is, we neglect
fermion and Higgs masses and the Higgs vacuum expectation value, the
effects of these being power-suppressed at high scales.  We
impose an infra-red cutoff $m_V$ on interactions that involve the
emission of an electroweak vector boson, $V=W^i$ for SU(2) or $B$
for U(1).\footnote{The cutoff is not strictly necessary for $B$
  emission, but we keep it because the $B$ and $W^3$ are mixed in the
  physical $Z^0$ boson.}
  Leading-order evolution kernels and one-loop running
couplings are used.  All the electroweak FFs
are generated dynamically by evolving upwards 
from a scale $q_0\sim m_V$.  In practice we take
 $q_0=m_V=100$ GeV. More details of the input FFs will be given in
 Section~\ref{sec:implementation}.

\subsection{Definition of the fragmentation functions}

The fragmentation function $D^k_i(x,q)$ gives the distribution of the
momentum fraction $x$ for particle species $k$ in a jet initiated by a
parton of type $i$ produced in a hard process at momentum scale $q$.
As in the case of PDFs, it is convenient to define the momentum-weighted FFs,
\beq
d_i^k(x,q) = \,x\,D_i^k(x,q)\,.
\eeq
Note that when we omit one of the labels $i$ or $k$, our expressions apply independent 
of its value.
One important thing to realize is that only particles in the
broken-symmetry phase (or the products of their decay or
hadronization) can be observed with a given momentum in the
detector, and the index $k$ therefore only runs over the particles in the broken
basis, that is, the fermions, photon, gluon, Higgs, $W^\pm$ and $Z^0$
bosons.  Furthermore, one typically does not distinguish between left-
and right-handed particles, or the different polarizations of the
vector bosons, in a detector. Thus, the total number of fermions is 6 quarks 
and anti-quarks, and 6 leptons and anti-leptons, giving 24
fermions. There are a total of 5 vector bosons and one Higgs, giving a
total of 30 particles we need to consider for $k$.

Since the index $i$ denotes the object produced at a high scale that
initiates the jet, we define it in the unbroken-symmetry phase.  When
$i$ is a fermion, one needs to separate left- and right-handed
chirality states,
which evolve differently as they belong to different representations
of the  SU(2) $\otimes$ U(1) symmetry.
This gives a total of 12 quarks and anti-quarks, and 9 leptons and
anti-leptons,  making 42 fermions. 
 
For each transversely-polarized SM vector boson, we need separate
positive and negative helicity FFs, $d^k_{V_\pm}$, since boson polarization
is generated during evolution and transmitted to the
fermions.\footnote{The original version of~\cite{Bauer:2017isx} did
  not discuss the effects of polarized vector bosons, the importance
  of which for electroweak evolution was first pointed out
  in~\cite{Manohar:2018kfx,Fornal:2018znf}.}
Interference between different helicity states cancels upon azimuthal
integration of transverse momenta in successive parton splittings, so
there are no mixed-helicity boson FFs.

Since SU(3) is unbroken, we need only a single gluon FF of each
helicity for each fragmentation product, $d^k_{g_+}$ and $d^k_{g_-}$.
For the SU(2) $\otimes$ U(1) symmetry, there are 8
transversely-polarized gauge bosons ($W_\pm^+$, $W_\pm^-$, $W^3_\pm$ and $B_\pm$), 
For the neutral bosons $B$ and $W^3$, one also needs to take into
account the two mixed $BW_\pm$ FFs, representing the interference contribution
when $i$ could have been either of them. 
Thus there are a total of 12 gauge boson labels required. 
There are a total of 4 Higgs bosons $H^\pm$, $H^0$ and $\overline H^0$
in the unbroken phase, and no mixed neutral Higgs FFs contribute for the
processes we shall consider. This brings the total to 58, the same number as 
for PDFs, as summarized in Table \ref{tab:FFcounts}.

\begin{table}[h!]
\begin{center}
\begin{tabular}{|c|ccc|c|}
\hline
\diagbox{i}{k}  & $f_{\rm light}$ & $V$ & $H$& sum\\\hline
$f$ & $42 \times 24$ & $42 \times 5$ & $42$ & $42 \times 30$\\
$g_\pm$ & $2 \times 24$ & $2 \times 5$ & $2$ & $2 \times 30$ \\
$W_\pm^\pm$ & $4 \times 24$ & $4 \times 5$ & $4$ & $4 \times 30$ \\
$V_\pm^0$ & $6 \times 24$ & $6 \times 5$ &  $6$ & $6 \times 30$ \\
$H^\pm$ & $2 \times 24$ & $2 \times 5$ & $2$ & $2 \times 30$ \\
$H^0$ & $2 \times 24$ & $2 \times 5$ &  $2$ & $2 \times 30$ \\
\hline
sum & $58 \times 24$ & $58\times 5$ & $58$ & $58 \times 30$ \\
\hline
\end{tabular}
\end{center}
\caption{\label{tab:FFcounts}Total number of fragmentation functions required. For a given final-state particle $k$, one requires a total of 58 FFs, 
which is the same as the number of PDFs needed for the initial
state. Each object $i$ can fragment into 30 particles $k$ (the total
number of particles and antiparticles 
in the Standard Model). Thus, in general $58 \times 30 = 1740$ FFs are required.}
\end{table}

Instead of using the unbroken basis, where all particles have definite
quantum numbers under the SU(3) $\otimes$ SU(2) $\otimes$ U(1), one 
can also work in the broken basis, where instead of $H^0$ and
$\overline H^0$ one has the Higgs boson $h$ and the longitudinally-polarized
$Z^0$, and instead of the neutral gauge bosons $B$ and $W_3$, one
has the photon and transversely-polarized $Z^0$. 
In the latter case, one can construct the FFs for the photon, the
$Z^0$ and their mixed $\gamma Z$ state as transformations
of the FFs for the $B$, the $W_3$ and their mixed state.  This is
anyway necessary at the electroweak scale, below which the symmetry is
broken.  Using $A = c_W B + s_W W_3$ and $Z^0 = - s_W B + c_W W_3$,
the relation between FFs containing $i = \gamma, Z, \gamma Z$ and
those with $i = B, W_3, BW$ is
\beqn\label{eq:fgamz}
\left(
\begin{array}{c}
	d_{\gamma} \\
	d_Z \\
	d_{\gamma Z} \\
\end{array}
\right)
=\left(
\begin{array}{ccc}
	c_W^2 & s_W^2 & c_W s_W  \\
	s_W^2 & c_W^2 & -c_W s_W \\
	-2c_W s_W & 2c_W s_W & c_W^2-s_W^2 \\
\end{array}
\right)
\left(
\begin{array}{c}
	d_{B} \\
	d_{W_3} \\
	d_{BW} \\
\end{array}
\right) \,,
\eeqn
and thus
\beqn\label{eq:fgamzInv}
\left(
\begin{array}{c}
		d_{B} \\
	d_{W_3} \\
	d_{BW} \\
\end{array}
\right)
=\left(
\begin{array}{ccc}
	c_W^2 & s_W^2 & -c_W s_W  \\
	s_W^2 & c_W^2 & c_W s_W \\
	2c_W s_W & -2c_W s_W & c_W^2-s_W^2 \\
\end{array}
\right)
\left(
\begin{array}{c}
	d_{\gamma} \\
	d_Z \\
	d_{\gamma Z} \\
\end{array}
\right) \,,
\eeqn
where the weak mixing parameters are given by
\begin{align}
s_W &\equiv s_W(q) = \sqrt{\frac{\alpha_1(q)}{\alpha_1(q)+\alpha_2(q)}}\nn
c_W &\equiv c_W(q) = \sqrt{\frac{\alpha_2(q)}{\alpha_1(q)+\alpha_2(q)}}
\,.
\end{align}

Although the flavor basis chosen above is the most intuitive, the
fact that many of the 58 FFs are coupled to one another makes it quite
difficult to solve the evolution equations. To decouple some of the
equations, it helps to change the basis such that the ingredients have
definite total isospin $\mathbf T$ and $\mathrm{CP}$ quantum numbers,
which (neglecting the tiny $\mathrm{CP}$ violation) are conserved in the
Standard Model. Then only FFs with the same quantum numbers can mix.
The FFs for each set of quantum numbers required are shown in
Table~\ref{tab:T_CPStates}.  In the case of the vector bosons, the
unpolarized FFs $d^k_{V_+}+d^k_{V_-}$ can have $\{\mathbf T, \mathrm{CP}\}
=\{0,+\},\{1,-\}$ or $\{2,+\}$, while the helicity asymmetries
$d^k_{V_+}-d^k_{V_-}$ have $\{0,-\},\{1,+\}$ or $\{2,-\}$.  Further
details of the isospin and CP basis are given in Appendix~\ref{app:isospin}.
\begin{table}[h!]
\begin{center}
\begin{tabular}{l|l}
$\{\mathbf T, \mathrm{CP}\}$ & fields\\\hline
$\{0,  \pm\}$ & $2 n_g\times q_R\,, n_g\times \ell_R\,, n_g\times q_L\,, n_g\times \ell_L\,, g\,, W\,, B\,, H$  \\
$\{1,  \pm\}$ & $n_g\times q_L\,, n_g\times \ell_L\,, W\,,BW, H$ \\
$\{2, \pm\}$ & $W$ \\
\end{tabular}
\end{center}
\caption{\label{tab:T_CPStates}The 58 FFs required for the SM evolution can written in a basis with definite conserved quantum numbers. $2(5 n_g+4)$ FFs contribute to the $\{0, \pm\}$ states, $2(2 n_g+3)$ to each to the $\{1, \pm\}$ and 2 to the $\{2, \pm\}$, where $n_g = 3$ stands for number of generations. }
\end{table}

Note that in general there can be additional mixed FFs, which however
are zero in our matching conditions at scale $q_0$ and are not generated in the
evolution. In particular, there can be states mixing left-and
right-handed fermions, but they are not present when we consider
only the FFs $d_i^k$ for unpolarized particles $k$.

\subsection{General evolution equations}
\label{sec:General_Evolution}
We consider the $x$-weighted FFs of parton species $i$
at momentum fraction $x$ and scale $q$, $d_i(x,q)$.
In leading order they satisfy evolution equations of the following
approximate form:\footnote{In Section~\ref{sec:leading_log_2} we
  present a modification of the evolution equations to achieve full
  leading-logarithmic accuracy.} 
\beqn
\label{eq:genevol}
q\frac{\pd}{\pd q} d^k_i(x, q) &=& \sum_I  \frac{\alpha_{I}(q)}{\pi} \left[  P^V_{i,I}(q) \, d^k_i(x, q) +  \sum_j  C_{ji,I} 
\int_x^{z_{\rm max}^{ji,I}(q)} \!\!\! \df z \, P^R_{ji, I}(z) d^k_j(x/z, q) \right] \nn
&\equiv& \sum_I \left[q\frac{\pd}{\pd q}  d^k_{i}(x, q)\right]_I
\,.
\eeqn
Here, the sum over $I$ goes over the different interactions in the
Standard Model
 and the notation $\left[q\,\pd / \pd q\, d^k_{i}(x, q)\right]_I$
 implies that we only keep the terms proportional to the coupling
 $\alpha_I$ when taking the derivative\footnote{Note that  $\left[
     \ldots \right]_I$ is only  introduced for notational convenience
   and should not be interpreted as setting all other couplings to
   zero. In particular, the FFs appearing on the right-hand side of
   \eq{genevol} still depend on the values of all coupling
   constants.}.
We denote by $I = 1, 2, 3$ the pure ${\rm U}(1)$, ${\rm SU}(2)$ and
${\rm SU}(3)$ gauge interactions, by $I =Y$ the Yukawa interactions, and 
by $I = M$ the mixed interaction proportional to
\beq
\alpha_M(q) = \sqrt{\alpha_1(q)\, \alpha_2(q)}
\,.\eeq
The first contribution in~\eq{genevol}, proportional to $P^V_{i,I}$, denotes the
virtual contribution to the FF evolution, while the second
contribution is the real contribution from the splitting of parton $i$
into parton $j$.  Notice that $i$ and $j$ are interchanged here
relative to the PDF evolution equations, because $d^k_j$ represents the
fragmentation of the outgoing parton from the splitting, rather than
the distribution of the incoming one.  The maximum value of $z$ in the
integration of the real contribution depends on the type of splitting
and interaction, and we choose
\beq
\label{eq:zmax}
z_{\rm max}^{ji,I}(q) = \Big\{
\begin{array}{ll}
1 - \frac{m_V}{q} & {\rm for}\, I = 1, 2, \,{\rm and}\, i, j \notin V \,{\rm or}\, i, j \in V
\\
1 & {\rm otherwise}
\end{array}
\,,\eeq
that is, we apply an infrared cutoff $m_V$, of the order of the
electroweak scale, when a $B$ or $W$ boson is emitted.  This regulates
the divergence of the splitting function for those emissions as $z\to
1$.  Such a cutoff is mandatory for $I=2$ because there are FF
contributions that are SU(2) non-singlets.  The evolution equations
for SU(3) are regular in the absence of a cutoff, as hadron FFs are
color singlets.  Similarly for U(1), the unpolarized FFs have zero
hypercharge,\footnote{Although there can be contributions with
  non-zero hypercharge for transversely polarized
  fermions~\cite{Ciafaloni:2005fm}.}  but we include the same cutoff
for $I = 1$, since the $B$ and $W_3$ are mixed in the physical $Z$ and
$\gamma$ states.

It was shown in~\cite{Bauer:2017isx} that the virtual corrections for
the fermion, scalar and unmixed, unpolarized vector boson PDFs,
which are the same for the corresponding FFs, are given by
\beq
\label{eq:PVirtualDef}
P^V_{i,I}(q) = - \sum_{j} C_{ji,I} \int_0^{z_{\rm max}^{ji,I}(q)} \!\!\! z\,\df z\,P^R_{ji,I}(z)
 \,  {\rm for}\, i \neq \rm{BW}\,.
\eeq
The virtual corrections for the individual vector boson helicity
states are the same as the unpolarized ones.  For the mixed FF one has
\beq\label{eq:PVmixed}
P^V_{BW,1}(q) = \frac{1}{2} P^V_{B,1}(q)\,, \qquad P^V_{BW,2}(q) = \frac{1}{2} P^V_{W,2}(q)
\,,\eeq
while the virtual contribution for $i=BW$  is zero for the other
interactions.

Thus for the unmixed FFs we have simply
\beq
\label{eq:genevol2}
\left[q\frac{\pd}{\pd q} d^k_i(x, q)\right]_I =\frac{\alpha_{I}(q)}{\pi} \sum_j  C_{ji,I} 
\int_0^{z_{\rm max}^{ji,I}(q)} \!\!\! \df z \, P^R_{ji, I}(z)
\left[d^k_j(x/z, q) -z\, d^k_i(x, q)\right].
\eeq
This implies that the DGLAP equations are defined by the splitting functions $P^R_{ji, I}(z)$ and the coefficients 
$C_{ji,I}$.

Most of the splitting functions can be obtained from the seminal paper of 
Altarelli and Parisi~\cite{Altarelli:1977zs}.  For the
gauge interactions of the Standard Model, $I=1,2,3$ and the mixed
interaction $M$, which we denoted collectively by $I =G$, one finds
\beqn
P^R_{f_Lf_L,G}(z) &=& P^R_{f_Rf_R,G}(z) =\frac 2{1-z}-(1+z) \,, \\
P^R_{V_+f_L,G}(z) &=& P^R_{V_-f_R,G}(z)=\frac{(1-z)^2}z\,,\\
P^R_{V_-f_L,G}(z) &=& P^R_{V_+f_R,G}(z)=\frac 1z\,,\\
P^R_{f_LV_+,G}(z) &=& P^R_{f_RV_-,G}(z) = \frac{1}{2} (1-z)^2\,,\\
P^R_{f_LV_-,G}(z) &=& P^R_{f_RV_+,G}(z) = \frac{1}{2} z^2\,,\\
P^R_{V_+V_+,G}(z) &=& P^R_{V_-V_-,G}(z) = \frac 2{1-z}+\frac 1z -1 -z(1+z)\,,\\
P^R_{V_+V_-,G}(z) &=& P^R_{V_-V_+,G}(z) = \frac{(1-z)^3}z\,,\\
P^R_{HH,G}(z) &=& \frac 2{1-z}-2\,,\\
P^R_{V_\pm H,G}(z) &=& \frac 1z -1\,,\\
P^R_{HV_\pm,G}(z) &=& \frac 12 z(1-z)\,.
\eeqn
The factor of $1/2$ in $P_{fV}$ has to be included since we are
considering fermions with definite chirality.  Notice also that we
have for splitting to and from antifermions, from CP invariance,
\beqn
&&P^R_{\bar f_LV_+,G}(z) = P^R_{f_LV_-,G}(z)\,,\;\;
P^R_{\bar f_RV_+,G}(z) = P^R_{f_RV_-,G}(z)\,,\\
&&P^R_{V_+\bar f_L,G}(z) = P^R_{V_-f_L,G}(z)\,,\;\;
P^R_{V_+\bar f_R,G}(z) = P^R_{V_-f_R,G}(z)\,.
\eeqn

Finally for the Yukawa interaction ($Y$), one has
\beqn
P^R_{ff,Y}(z) &=& \frac{1-z}{2} \,, \\
P^R_{Hf,Y}(z) &=& P^R_{ff,Y}(1-z)\,,\\
P^R_{fH,Y}(z) &=& \frac{1}{2}\,.
\eeqn
We now give the necessary coefficients $C_{ij,I}$ for the five interactions.
\\
\\
{\bf I=3: SU(3) interaction}\\
We start by considering the well known case of SU(3) interactions. The
relevant degrees of freedom are the gluon, as well as left and
right-handed quarks. The coupling coefficients are
\beq
\label{eq:SU3Couplings}
C_{qq,3} = C_{gq,3} = C_F\,, \qquad C_{qg,3} = T_R\,, \qquad C_{gg,3} = C_A
\,,
\eeq
where $C_F = 4/3$, $C_A = 3$, $T_R = 1/2$. Note that since SU(3) has the same 
coupling to left- and right-handed quarks, it does not produce a
polarization asymmetry on its own, unless an initial asymmetry is
present due to polarized initial states.  However, due to the
different electroweak evolution of the left- and right-handed
fermions, even the gluon FFs develop a polarization
asymmetry above the electroweak scale.
\\
\\
{\bf I=1: U(1) interaction}\\
For  ${\rm U}(1)$ the relevant degrees of freedom are left- and
right-handed fermions (denoted by the subscript $f$), as well as the
${\rm U}(1)$ gauge boson $B$. The coefficients involving fermions and
gauge bosons are
\beq
C_{ff,1} = C_{Bf,1} = Y_f^2\,, \qquad C_{fB,1} = N_f \, Y_f^2\,, \qquad C_{BB,1} = 0\,,
\eeq
where the hypercharges of the different fermions are given by $Y_{q_L} = 1/6$, $Y_{u_R} = 2/3$, $Y_{d_R} = -1/3$, $Y_{\ell_L} = -1/2$ and $Y_{e_R} = -1$. 
The color factor $N_f$ is equal to $N_c=3$ for quarks and 1 for
leptons. The coefficients involving the Higgs bosons are
\begin{align}
C_{hh,1} = C_{Bh,1} = C_{hB,1} = \frac{1}{4}
\,,
\end{align}
where $h$ here stands for any of the four Higgs boson FFs.
\\
\\
{\bf I=2: SU(2) interaction}
\\
Denoting by $u_L$ and $d_L$ any
up- and down-type left-handed fermion, one finds
\beqn\label{eq:SU2_Cq}
C_{u_L d_L, 2} = C_{d_L u_L, 2} = C_{W^+ u_L, 2} = C_{W^- d_L, 2} &=& \frac 12\,,\\
C_{u_L u_L, 2} = C_{W_3 u_L, 2} = C_{d_L d_L, 2} = C_{W_3 d_L, 2}  &=& \frac 14\,,\\
C_{u_L W^+, 2}=C_{d_L W^-, 2} &=& N_f\frac 12\,,\\
C_{u_L W_3, 2} = C_{d_L W_3, 2} &=& N_f\frac 14\,,\\
\label{eq:SU2_CW}
C_{W^\pm W^\pm, 2} = C_{W^\pm W_3, 2} = C_{W_3 W^\pm, 2} &=& 1\,,
\eeqn
where as before the color factor $N_f=3$ for quarks, 1 for leptons.
The coupling coefficients of the Higgs bosons are given by
\beqn
C_{H^+ H^+,2} &=& C_{H^0 H^0,2} = C_{W_3 H^+,2} =  C_{W_3 H^0,2}\,,\nn
&=& C_{H^+ W_3,2} = C_{H^0 W_3,2} = \frac{1}{4}\,,\\
C_{H^+ H^0,2} &=& C_{H^0 H^+,2} = C_{H^+ W^+,2} = C_{W^+ H^+,2}\,,\nn
&=& C_{H^0 W^-,2} = C_{W^- H^0,2} = \frac{1}{2}
\,.
\eeqn
The couplings for the charge-conjugate states are the same.
\\
\\
{\bf I=Y: Yukawa interaction}
\\
In this work we only keep the top Yukawa coupling, setting all others to zero. This gives the following coefficients:
\beq
C_{q_L^3 t_R,Y} = C_{H^0 t_R, Y} = C_{H^+ t_R, Y} = C_{t_R q_L^3, Y} = C_{\bar H^0 t_L, Y} = C_{H^- b_L, Y} = 1
\,,
\eeq
where $q_L^3$ denotes either the left-handed top or bottom quark. We furthermore need
\beq
C_{t^R H^0, Y} = C_{t^R H^+, Y} = C_{t^L \bar H^0, Y} = C_{b_L H^-, Y} =  N_c
\,.
\eeq
\\
{\bf\boldmath I=M: Mixed $B-W_3$ interaction}
\\
Finally, we need to consider the evolution involving the mixed $BW$
boson FF.  The non-vanishing couplings are
\beqn
C_{BWf_u,M} = -C_{BWf_d,M} &=& 2\frac{Y_f}{2}\,,\\
C_{f_uBW,M} = - C_{f_dBW,M} &=& N_f \frac{Y_f}{2}\,,
\eeqn
where $f_u$ and $f_d$ represent the  up- and down-type left-handed fermions
and anti-fermions of all generations. Since $Y_{\bar f}=-Y_f$ and $T_{3\bar f}=-T_{3f}$, the
couplings for fermions and anti-fermions are identical.
 The factor of 2 in the first line comes from our definition of
$f_{BW}$ as the sum of $BW$ and $WB$ contributions.
The diagonal coefficients $C_{f_uf_u,M}$ and $C_{f_df_d,M}$ are zero
because  there is no vector boson with both U(1) and SU(2)
interactions.  The couplings involving the Higgs bosons are
 \beqn
C_{BW H^+,M} =  -C_{BW H^0,M} &=& \frac 12\,,\\
C_{H^+ BW,M} = -C_{H^0 BW,M} &=& \frac{1}{4}\,,
\eeqn
 where, as for the fermions, the same relations hold for the
charge-conjugate states.

The resulting evolution equations in the  $\{\mathbf T, \mathrm{CP}\}$
basis are given in full in Appendix~\ref{app:forward}.

\subsection{Double logarithmic evolution}
\label{sec:double_log}
Any combination of FFs that is not SU(2)-symmetric has a component
that evolves double-logarithmically.  For example, from
Eqs.~(\ref{eq:genevol2}) and (\ref{eq:SU2_Cq}-\ref{eq:SU2_CW}),
the combination of left-handed quark FFs that has $\{\mathbf T,
\mathrm{CP}\} = \{1,-\}$,
\beq
d^{1-}_q=\frac 14 \left(d_{u_L}-d_{d_L}-d_{\bar u_L}+d_{\bar d_L}\right),
\eeq
 satisfies the evolution equation
\beqn\label{eq:evol_d1}
\left[q\frac{\pd}{\pd q} d^{1-}_q(x, q)\right]_2 &=&\frac{\alpha_{2}(q)}{\pi}\biggl\{
\int_0^{1-m_V/q} \!\!\! \df z \, P^R_{ff, G}(z)\left[-\frac 14
  d^{1-}_q(x/z, q) -\frac 34 z\, d^{1-}_q(x,
  q)\right]\nn
&&+\int_0^1 \df z \, P^R_{Vf, G}(z)\left[\frac 12
  d^{1-}_W(x/z, q) -\frac 34 z\, d^{1-}_q(x,
  q)\right]\Biggr\}
  \,,
\eeqn
where
\beq
d^{1-}_W=\frac 12 \left(d_{W^+_+}-d_{W^-_+}+d_{W^+_-}-d_{W^-_-}\right).
\eeq
The mismatch between the
coefficients of $d^{1-}_q(x/z,q)$ and $d^{1-}_q(x,q)$ on
the right-hand side of~\eq{evol_d1} induces a logarithmic sensitivity to $m_V$ and
hence a double-logarithmic term in the evolution.  In fact, noting that
the SU(2) contribution to the fermion Sudakov factor is
\beqn
\Delta_{f,2}(q)&=&\exp\left\{-\frac 34\int_{m_V}^q\frac{\df q'}{q'}\frac{\a_2(q')}{\pi}\left[\int_0^{1-m_V/q'} \!\!\!
\df z \,z\, P^R_{ff, G}(z)+\int_0^1 \df z \,z\, P^R_{Vf,
  G}(z)\right]\right\}\nn
&=&\exp\left\{-\frac 34\int_{m_V}^q\frac{\df q'}{q'}\frac{\a_2(q')}{\pi}\left[\int_0^{1-m_V/q'} \!\!\!
\df z \, P^R_{ff, G}(z) +{\cal O}(m_V/q')\right]\right\},
\eeqn
we have
\beqn
&&\left[q\frac{\pd}{\pd q} d^{1-}_q(x, q)\right]_2 =\frac{\alpha_{2}(q)}{\pi}\Biggl\{
-\frac 14\int_0^1 \df z \, P^R_{ff, G}(z)\left[
  d^{1-}_q(x/z, q) -d^{1-}_q(x,
  q)\right]\nn
&&\;\;+\frac 12\int_0^1 \df z \, P^R_{Vf, G}(z)  d^{1-}_W(x/z, q)
+\frac 43 d^{1-}_q(x,q)\,q\frac{\df}{\df q}\ln\Delta_{f,2}(q)+ {\cal O}(m_V/q)\Biggr\}.
\eeqn
The  integrals are now independent of $m_V$ and therefore only produce
single-logarithmic evolution.  All the double-logarithmic dependence
comes from the Sudakov factor and we can write
\beq
d^{1-}_q(x, q) = \tilde d^{1-}_q(x, q)
\left[\Delta_{f,2}(q)\right]^{4/3}
\eeq
where $\tilde d^{1-}_q$ has only single-logarithmic evolution.
Similarly, all other FF combinations that are not SU(2)-symmetric
are suppressed at high energy by powers of the corresponding SU(2)
Sudakov factor~\cite{Ciafaloni:2005fm}. 

While for fermions there are only isospin 0 and 1 combinations
possible, for vector bosons one can also form combinations with
${\mathbf T}=2$:
\begin{align}
d^{2\pm}_W = \frac 16\left[\left(d_{W^+_+}+d_{W^-_+}-2d_{W^3_+}\right)
\pm \left(d_{W^+_-}+d_{W^-_-}-2d_{W^3_-}\right)\right].
\end{align}
The  double-logarithmic dependence in fact only depends on the value
of the isospin, and in general one finds
\beq
d^{T\pm}_i(x, q) = \tilde d^{T\pm}_i(x, q) \, \Delta_i^{(T)}(q)
\eeq
where in double-logarithmic approximation
\beq\label{eq:DeltaI}
\Delta_i^{(T)}(q)\simeq\exp\left[-T(T+1)\int_{m_V}^q\frac{\df q'}{q'}\frac{\a_2}{\pi}\int_0^{1-m_V/q'} \!\!\!
\frac{\df z}{1-z}\right]=
\exp\left[-T(T+1)\frac{\a_2}{2\pi}\ln^2\left(\frac q{m_V}\right)\right].
\eeq

\subsection{Momentum conservation}
\label{sec:mom_conservation}

The total momentum fraction carried by particle species $k$
in a jet initiated by a parton of type $i$ at scale $q$ is given by
\beq
\label{eq:momFracDef}
\vev{d_i^k(q)}\equiv \int_0^1 \df x\,d_i^k(x,q)\,.
\eeq
Noting that
\beq
\int_0^1 \df x\,P^R_{ji, I}\otimes d^k_j =Q_{ji,I}(q)\,\vev{d_j^k(q)} \,,
\eeq
where
\beq\label{eq:Qjidef}
Q_{ji,I}(q) = \int_0^{z_{\rm max}^{ji,I}(q)} \!\!\! \df z\,
z\,P^R_{ji, I}(z)\,,
\eeq
we have from the evolution equation~\eqref{eq:genevol2} for unmixed FFs
\beq
\label{eq:MomentumConservation}
\left[ q\frac{\df}{\df q}\vev{d^k_{i}(q)}\right]_I =\frac{\alpha_{I}(q)}{\pi} 
\sum_j C_{ji,I} Q_{ji,I}(q)\left[\vev{d^k_j(q)}  -\vev{d^k_i(q)}\right].
\eeq
Writing
\beq
F_{ij}(q) = \sum_I\frac{\alpha_{I}(q)}{\pi} \left[
C_{ji,I} Q_{ji,I}(q) -\delta_{ij}\sum_l C_{li,I} Q_{li,I}(q)\right]
\eeq
this gives
\beq
q\frac{\df}{\df q}\vev{d^k_{i}(q)} = \sum_j F_{ij}(q)\,\vev{d^k_j(q)}\,.
\eeq
This is a set of ordinary differential equations that can be solved by finding
the eigenvalues and eigenvectors of the matrix $F_{ij}(q)$.  One of
the eigenvalues, corresponding to the eigenvector $(1,1,\ldots,1)$, is
zero, so for every species $k$ and unmixed interaction $I$ there is a
linear combination of the momentum fractions $\vev{d^k_i}$ that is
scale-independent.  Furthermore, since the sum of momenta of all
species $k$ in the jet must equal that of the initial parton $i$, for
the unmixed FFs we have
\beq\label{eq:mom_cons}
\sum_k\vev{d^k_i(q)} = 1 
\eeq 
 for every value of $i$, and thus
\beq
\left[ q\frac{\df}{\df q}\sum_k\vev{d^k_{i}(q)}\right]_I =0\,,
\eeq
so the momentum sum is conserved by each interaction separately.

For the mixed FF, $i=BW$, of either helicity, the real emission term involves the
difference between the momentum sums for up- and down-type fermions
and scalars, which vanishes, so that, from~\eq{PVmixed},
\beq
q\frac{\df}{\df q} \sum_k\vev{d^k_{BW}} =\frac 1{2\pi}\left[\a_1(q)P^V_{B,1}(q)+\a_2(q)P^V_{W,2}(q)\right]\sum_k\vev{d^k_{BW}}
\eeq
and hence
\beqn
\sum_k\vev{d^k_{BW}(q)} &=& \exp\left\{\int_{q_0}^q\frac
  1{2\pi}\frac{\df
    q'}{q'}\left[\a_1(q')P^V_{B,1}(q')+\a_2(q')P^V_{W,2}(q')\right]\right\}\sum_k\vev{d^k_{BW}(q_0)}\nn
&=& \Delta_{BW}(q)\sum_k\vev{d^k_{BW}(q_0)}
\,,
\eeqn
where $\Delta_{BW}(q)$ is the $BW$ Sudakov factor.  Now from~\eq{fgamzInv} we have
\beqn
\sum_k \vev{d^k_{BW}} &=& \sum_k\left[2c_Ws_W\left(\vev{d^k_\gamma}
  -\vev{d^k_Z}\right)+\left(c_W^2-s_W^2\right)\vev{d^k_{\gamma Z}}\right]\nn
&=&\left(c_W^2-s_W^2\right)\sum_k\vev{d^k_{\gamma Z}}
\,,
\eeqn
and, as will be discussed in Section~\ref{sec:implementation}, the
mixed $\gamma Z$ FFs $d^k_{\gamma Z}$ all vanish at the electroweak
scale $q=q_0$. Hence the momentum sum for the mixed FF, of either helicity, vanishes at all
scales:
\beq\label{eq:mom_sum_BW}
\sum_k \vev{d^k_{BW}(q)}\equiv 0\,.
\eeq

\section{Achieving full (next-to-) leading-logarithmic accuracy}
\label{sec:leading_log_2}
We have seen in Section \ref{sec:double_log} that fragmentation
functions that are not iso-singlets experience double-logarithmic
evolution. This is due to the fact that the soft singularity as $z \to
1$ in the splitting functions $P^R_{ii,G}(z)$ do not cancel between
the virtual and real contributions. This is the origin of the SU(2)
Sudakov factor, which according to~\eq{PVirtualDef} is given by
\beqn\label{eq:Deltai2}
\Delta_{i,2}(q) &=& \exp\left[\int_{m_V}^q\frac{\df
    q'}{q'}\frac{\a_2(q')}{\pi} P^V_{i,2}(q')\right]\nn
 &=& \exp\left[-\int_{m_V}^q\frac{\df
    q'}{q'}\frac{\a_2(q')}{\pi} \sum_{j} C_{ji,I} \int_0^{z_{\rm
      max}^{ji,I}(q')} \!\!\! z\,\df z\,P^R_{ji,I}(z)\right]
\,.
\eeqn
The leading logarithmic contribution arises from the term in the splitting function that is divergent as $z \to 1$ and one can write
\beq
C_{ji,I} \int_0^{z_{\rm
      max}^{ji,I}(q')} \!\!\! z\,\df z\,P^R_{ji,I}(z)\sim
2\,C_{i,2}\int_0^{1-m_V/q'}\frac{\df z}{1-z}
=2\,C_{i,2}\ln\left(\frac{q'}{m_V}\right),
\eeq
where $C_{f,2}=C_{H,2}=3/4$ and $C_{W,2}=2$.
For a fixed coupling $\a_2$ we then obtain the double-logarithmic (DL)
approximation to the Sudakov factor,
\beq
\Delta^{DL}_{i,2}(q) = \exp\left[-C_{i,2}\frac{\a_2}{\pi}
  \ln^2\left(\frac
    q{m_V}\right)\right]\equiv\exp\left[-\frac{C_{i,2}}{\pi}\a_2 L^2\right].
\eeq

However, it is well known that in general, Sudakov factors take the form
\beq\label{eq:SudGeneral}
\Delta_{i,2}(q)=\exp\left[L\,g_1(\a L)+g_2(\a L)+\a \,g_3(\a L)+\ldots\right]
\eeq
where in the case at hand $\a \equiv\a_2(q)$.  The functions $g_i$ determine the logarithmic terms necessary 
in the expansion when the size of the log is such that $\a L\sim 1$. In this case, the function $g_1$ sums all leading logarithms
(LL), $g_2$ sums next-to-leading
logs (NLL), and so on.  The double logarithmic expansion, on the other
hand, is sufficient if the size of the log satisfies $\alpha L^2 \sim
1$ but $\alpha L \ll 1$. 

At the highest energies reachable at the LHC and a future 100 TeV collider the logarithm can be as large 5 and 7, respectively. Given that
$\a_2\sim 0.03$, this means that $\a_2 L^2 \sim 1$, but one still has $\a_2 L \ll 1$. This explains why the double logarithmic resummation 
we considered in~\cite{Bauer:2017isx,Bauer:2017bnh} and so far in this paper should be sufficient phenomenologically. In Table~\ref{tab:scaling} we show the dominant term missed when using DL, LL and NLL  resummation at partonic center-of-mass energies of $q \sim 1, 5, 30$ TeV. For each we also give the size of the first missed term if the resummed results are matched to the full ${\cal O}(\a)$ calculation. 
\begin{table}[h!]
\begin{center}
\begin{tabular}{l|cc|cc|cc|}
& \multicolumn{2}{c|}{DL} & \multicolumn{2}{c|}{LL} & \multicolumn{2}{c|}{NLL}\\
& no match & match & no match & match &no match & match \\\hline
missed term & $\a_2 L$ & $\a_2^2 L^3$ & $\a_2 L$ & $\a_2^2 L^3$ & $\a_2$ & $\a_2^2 L^2$\\\hline
$q \sim  1$ TeV & 0.08 & 0.02 & 0.08 & 0.02 & 0.03 & 0.006  \\
$q \sim  5$ TeV & 0.12 & 0.06 & 0.12 & 0.06 & 0.03 & 0.02 \\
$q \sim  30$ TeV & 0.18 & 0.19 & 0.18 & 0.19 & 0.03
                   & 0.03 \\ \hline
\end{tabular}
\end{center}
\caption{\label{tab:scaling} {Scaling of the dominant missed term in the perturbative expansion, for 
the double log
expansion, where only the leading $\alpha L^2$ term in the exponent is
kept, the LL expansion, where the whole function $g_1(\alpha L)$ is
kept, and the NLL expansion, where the functions $g_1(\alpha L)$ and
$g_2(\alpha L)$ are kept. For each of these, we show the scaling of
the first missed term if just the logarithmic resummation is used, and
also the scaling if the resummed result is matched with the fixed
order NLO calculation (such that the full $\alpha$ dependence is
reproduced).
}}
\end{table}
One can see that up to scales of order $1\, {\rm TeV}$ the double
logarithmic expansion is sufficient to give ${\cal O}(1\%)$ accuracy
as long as it is matched with the fixed order calculation at NLO, as
described in~\cite{Bauer:2017bnh}. Adding the full LL resummation
[using the complete function $g_1(\alpha L)$] does not improve the
situation, since one is still missing a term of order $\alpha^2 L^3$,
which comes from the missed $\alpha L$ term of the function
$g_2(\alpha L)$ multiplying the $\alpha L^2$ term of $Lg_1(\alpha
L)$. This term is only reproduced once the complete NLL resummation is
taken into account. This of course makes sense, since the full LL
resummation is only formally improving the accuracy of the DL
resummation when counting $\alpha L \sim 1$. In that limit, however,
the NLL resummation provides an ${\cal O}(1)$ effect, which needs to
be included as well. Note that the two different choices for the
scaling of the logarithm were already discussed in some detail
in~\cite{Chiu:2008vv}. 

Even though the full LL resummation does not improve the situation
over matched DL resummation for feasible collider energies, we will show
how it can be obtained in the DGLAP formalism by choosing the scale of the
running SU(2) coupling appropriately. It is well known in standard QCD
resummation and parton shower algorithms, that for double
logarithmically sensitive observables the evolution should be
angular-ordered and the running coupling should be evaluated at the
transverse momentum of gauge boson
emission~\cite{Dokshitzer:1978hw,Amati:1980ch}.  This means that instead of using
$\alpha_2(q)$ as we have been doing in the DGLAP evolution, one should
use $\alpha_2(q(1-z))$.  Then since
\beq\label{eq:a2run}
\a_2(q') = \frac{\a_2(q)}{1+\beta^{(2)}_0\frac{\a_2(q)}{\pi}\ln\frac{q'}{q}}
\,,
\eeq
with $\beta^{(2)}_0=19/12$,
the ratio of these two scale choices is given by the expansion
\begin{align}
\frac{\alpha_2(q(1-z))}{\alpha_2(q)} = 1 - \frac{\alpha_2(q)}{\pi} \beta_0^{(2)} \ln (1-z) + \left[\frac{\alpha_2(q)}{\pi} \beta_0^{(2)} \ln (1-z)\right]^2 + \ldots \,.
\end{align}
Note that these logarithmic terms in $1-z$ only give rise to large
logarithms if integrated against a singular function $f(z) \sim 1 /
(1-z)$. Thus, in standard DGLAP evolution in QCD, where the soft
divergence as $z \to 1$ cancels between the virtual and real
contributions, the difference between these two scales do not lead to
logarithmic terms that need to be resummed. For the case of SU(2)
DGLAP evolution of PDFs or FFs that are not iso-singlets, however,
this cancelation does not happen, and one finds
\begin{align}
\int_0^{1-\frac{m}{q}} \df z \frac{\alpha_2(q(1-z))}{\pi} \frac{1}{1-z} = \frac{\alpha_2(q)}{\pi} L +  \frac{\alpha_2^2(q)}{\pi^2} \frac{\beta_0^{(2)}}{2} L^2 + \ldots
\,,
\end{align}
which generates the LL function $g_1(\a_2 L)$.  The full LL
resummation is therefore obtained by changing the SU(2) splitting functions
that are singular as $z \to 1$ as
\begin{align}
\label{eq:Pff2}
P^R_{ff,2}(z) & \to P^R_{ff,2}(z, q) = \frac{\alpha_2[q(1-z)]}{\alpha_2(q)}\frac{2}{1-z} - (1+z)\,,
\\
\label{eq:PVV2}
P^R_{V_+V_+,2}(z) & \to P^R_{VV,2}(z, q) = \frac{\alpha_2[q(1-z)]}{\alpha_2(q)}\frac{2}{1-z} + \frac{1}{z} - 1 - z(1+z)\,,\\
\label{eq:PHH2}
P^R_{HH,G}(z) & \to P^R_{HH,G}(z, q) = \frac{\alpha_2[q(1-z)]}{\alpha_2(q)}\frac{2}{1-z} - 2\,.
\end{align}

By making one more change one can in fact also reproduce the full NLL
resummation of the collinear evolution. The only missing term is the
2-loop cusp anomalous dimension, which can be included using the CMW
prescription~\cite{Catani:1990rr} for the coupling constant. This
amounts to changing
\beq\label{eq:a2CMW}
\a_2[q(1-z)] \to \alpha^{\rm CMW}_2[q(1-z)]
\eeq
in Eqs.~(\ref{eq:Pff2}-\ref{eq:PHH2}), where
\begin{align}
\alpha^{\rm CMW}_2[q(1-z)] \equiv \alpha_2[q(1-z)]\left[ 1 +
  \frac{\Gamma_{\rm cusp, f}^{(2)}}{\Gamma_{\rm cusp, f}^{(1)}}
  \frac{\alpha_2[q(1-z)]}{\pi} \right]\simeq  \alpha_2[k_{\rm CMW}q(1-z)]\,,
\end{align}
\beq
k_{\rm CMW}=\exp\left(-\frac 1{\beta_0^{(2)}} \frac{\Gamma_{\rm cusp,
      f}^{(2)}}{\Gamma_{\rm cusp, f}^{(1)}}\right)
      \,,
\eeq
and $\Gamma_{\rm cusp, f}^{(n)}$ and $\Gamma_{\rm cusp, a}^{(n)}$
denote the cusp anomalous dimension in the fundamental and adjoint
representations at $n$-loop order.  For $n_g$ fermion generations and
$n_H$ Higgs doublets~\cite{Chiu:2007dg}
\beq
 \frac{\Gamma_{\rm cusp, f}^{(2)}}{\Gamma_{\rm cusp, f}^{(1)}} =
\frac{\Gamma_{\rm cusp, a}^{(2)}}{\Gamma_{\rm cusp, a}^{(1)}} = 
\frac{67}{18}-\frac{\pi^2}6-\frac 59 n_g-\frac 19 n_H = \frac{35}{18}-\frac{\pi^2}6\,,
\eeq
which gives
\beq
k_{\rm CMW} = \exp\left(\frac{6\pi^2-70}{57}\right) = 0.828\,.
\eeq
One can verify that this reproduces the complete NLL resummation in the collinear sector by comparing directly against the results of~\cite{Manohar:2018kfx}. For observables that are completely inclusive over the final state, where no soft function is required, this therefore reproduces the full NLL resummation. For less inclusive observables, it misses the logarithmic resummation coming from the evolution of the soft function, which was discussed in~\cite{Manohar:2018kfx} and is not included here.

As we have explained, including the full LL resummation, compared with only the DL resummation, does not improve the formal accuracy of the calculation, unless the full NLL effects are included at the same time. Nevertheless, we show its numerical effect when presenting results in Section~\ref{sec:results}.

\section{Implementation details and input FFs}
\label{sec:implementation}
For simplicity we start the evolution of all FFs at the electroweak breaking
scale $q_0\sim m_V$, which in practice we take to be 100 GeV.    Each
value of the fragmentation product $k$ requires a separate run of the
evolution code.  For a quark or charged lepton, $k=f$, assuming that the helicity of the
fragmentation product is not detected, we take as input
\beq
d_{f_L}^f(x,q_0)=d_{f_R}^f(x,q_0)=\delta(1-x)\,,
\eeq
setting all other initial FFs to zero.  Then the FFs for all 58 SM
states $i$ fragmenting into $f$ are generated by evolving
these input FFs to higher scales using the SM DGLAP equations given in
Section~\ref{sec:leading_log_2}.  
To obtain FFs at scales below $q_0$, the resulting FFs $d^f_i(x,q>q_0)$
should be convoluted with the $SU(3)\otimes U(1)_{\rm em}$-evolved and
hadronized FF of a jet of flavor $f$ produced at scale $q_0$.
The neutrinos $k=\nu$ have no right-handed states, so
the initial condition becomes
\beq
d_{\nu_L}^\nu(x,q_0)=\delta(1-x)\,,
\;\;\; d_i^\nu(x,q_0)=0\;\mbox{otherwise}\,.
\eeq
for evolution from scale $q_0$. The resulting FFs can be
interpreted directly as neutrino momentum fraction distributions,
since the neutrinos do not evolve below the electroweak scale.  

For fragmentation into a gauge boson $V$ we again assume the
helicity is not detected, so the input is
\beq
d^V_{V_+}(x,q_0)=d^V_{V_-}(x,q_0)=\delta(1-x)\,,
\;\;\; d_i^V(x,q_0)=0\;\mbox{otherwise}\,.
\eeq
For the gluon, the SM-evolved FFs at higher scales then
need to be convoluted with the FFs of a gluon jet produced at
scale $q_0$.  For the $W^\pm$, on the other hand, the 
boson can simply be allowed to decay at scale $q_0$.  For the neutral
gauge bosons $V=\gamma,\,Z^0$ we resolve them into the
unbroken $B,\,W^3$ and $BW$ states according to~\eq{fgamz} at
scale $q_0$ and evolve these upwards.  Again, the heavy bosons can
then decay directly at scale $q_0$, while the photon can either be
treated as a stable particle or fragmented by $U(1)_{\rm em}$ evolution at
lower scales.  Similarly the Higgs and longitudinal gauge boson FFs are
resolved as
\beqn
&&d^k_{W^+_L} = d^k_{H^+}\,,\;\; d^k_{W^-_L} = d^k_{H^-}\,,\\
&&d^k_{Z^0_L} = d^k_{h} =\frac 12\left(d^k_{H^0} + d^k_{\bar H^0}\right)\,,
\eeqn
and these are evolved to higher scales using the unbroken SM.

Notice that the momentum conservation relations (\ref{eq:mom_cons})
and (\ref{eq:mom_sum_BW})
involve sums over independent runs of the evolution code for
the 30 possible fragmentation products $k$, and must hold for each
one of the 58 fragmenting objects $i$, which provides a valuable check
on the correctness and precision of the code.

\section{Results}
\label{sec:results}
\FIGURE[h]{
 \centering
  \includegraphics[scale=0.42]{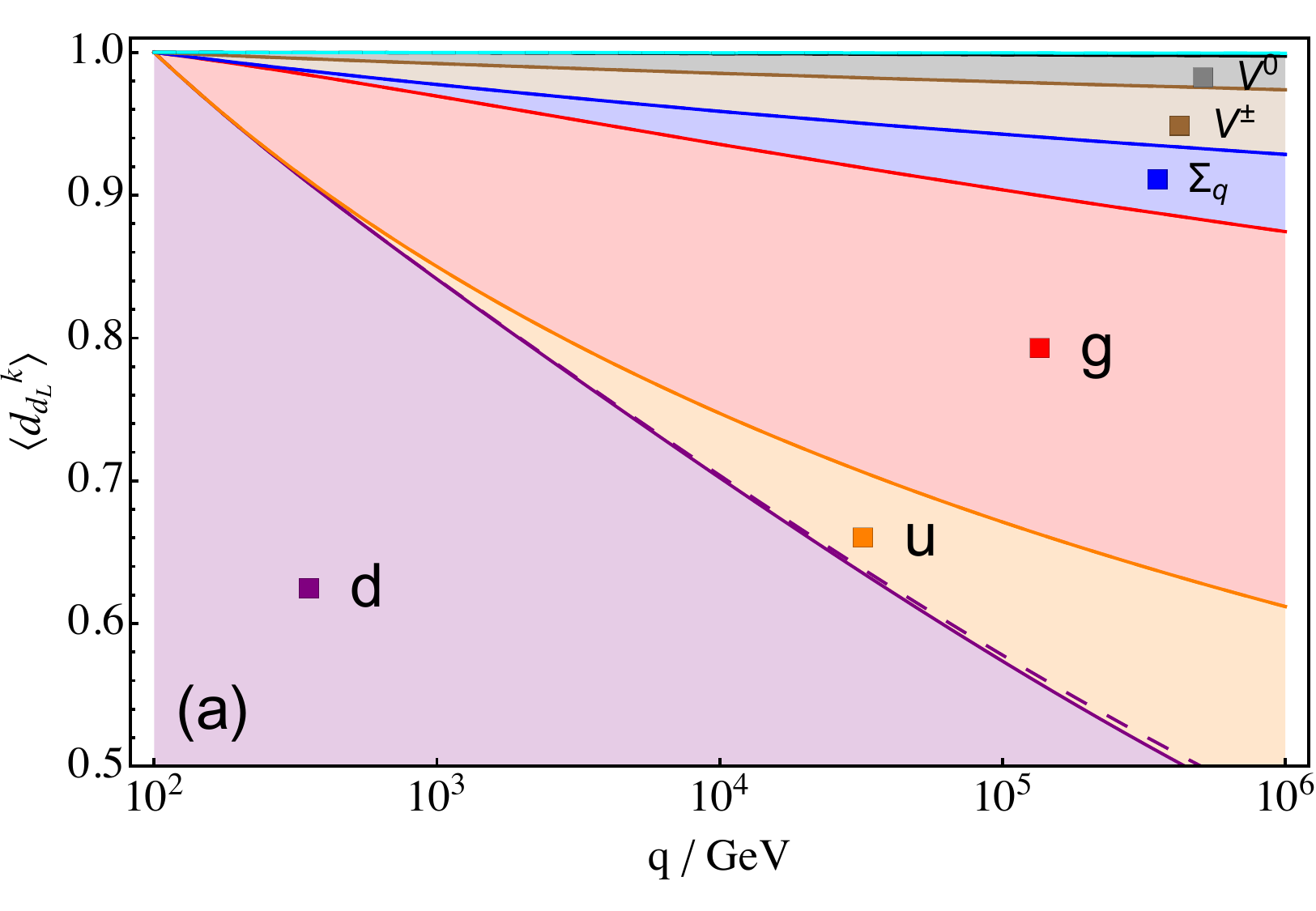}
  \includegraphics[scale=0.42]{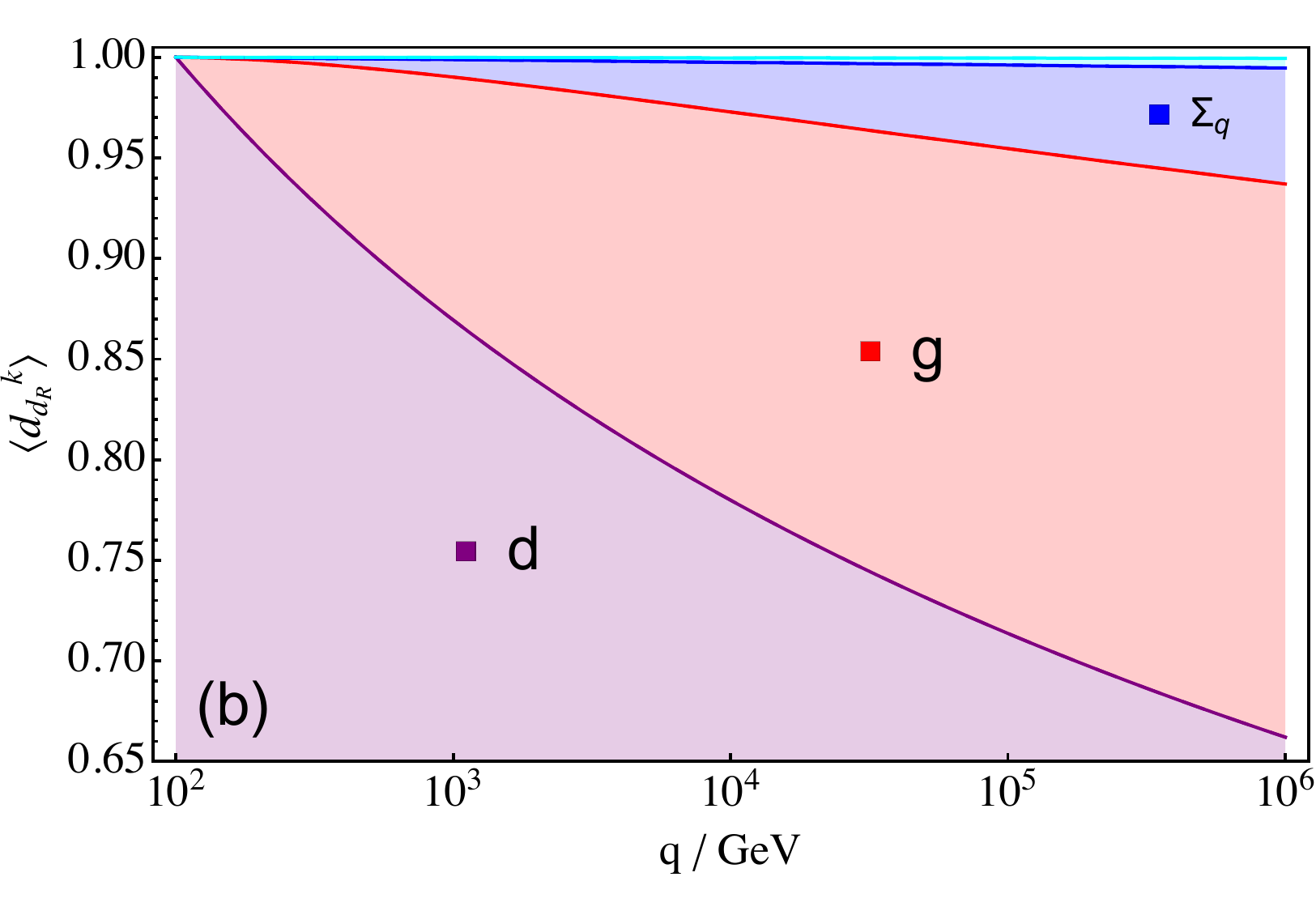}
  \includegraphics[scale=0.42]{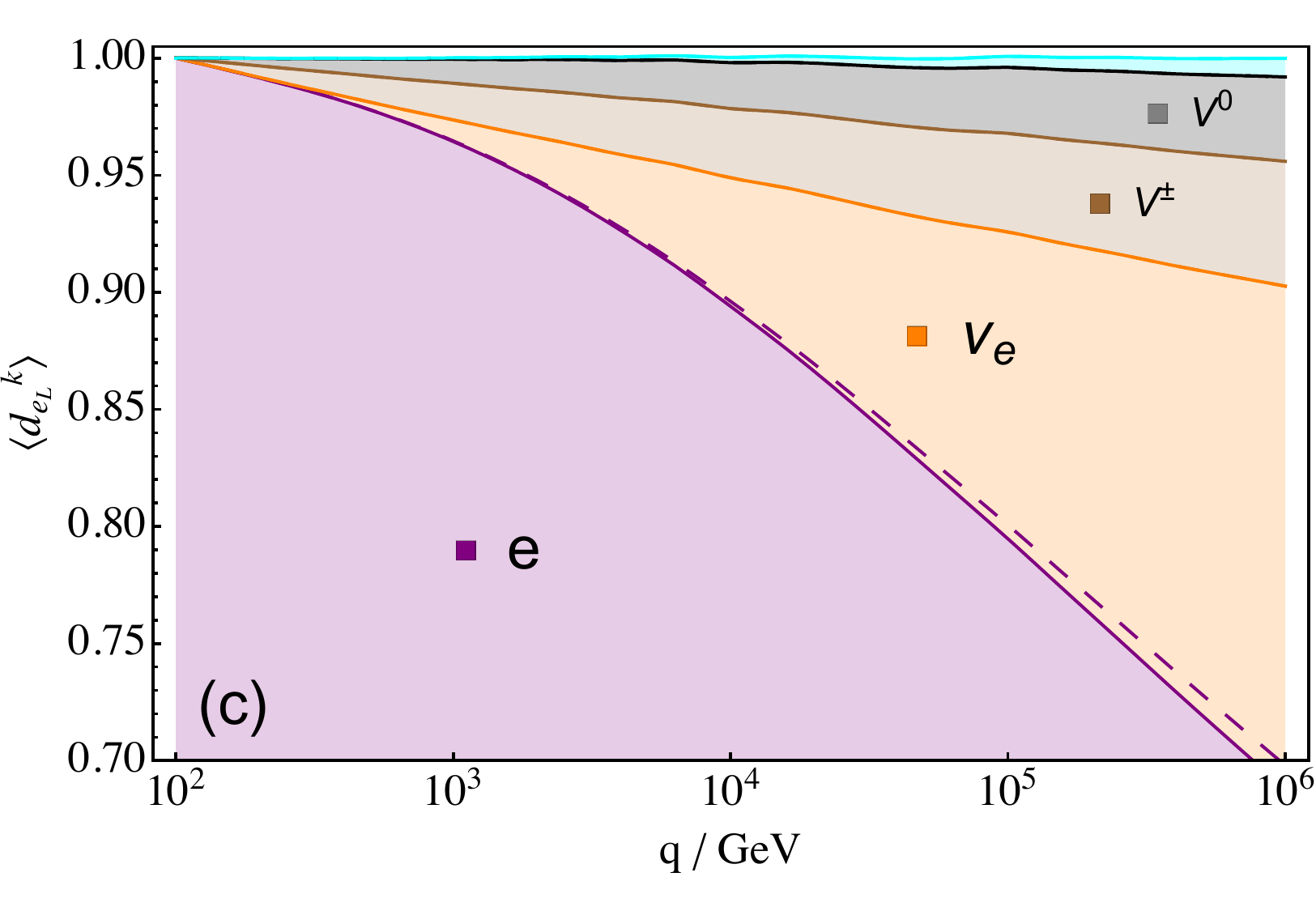}
  \includegraphics[scale=0.42]{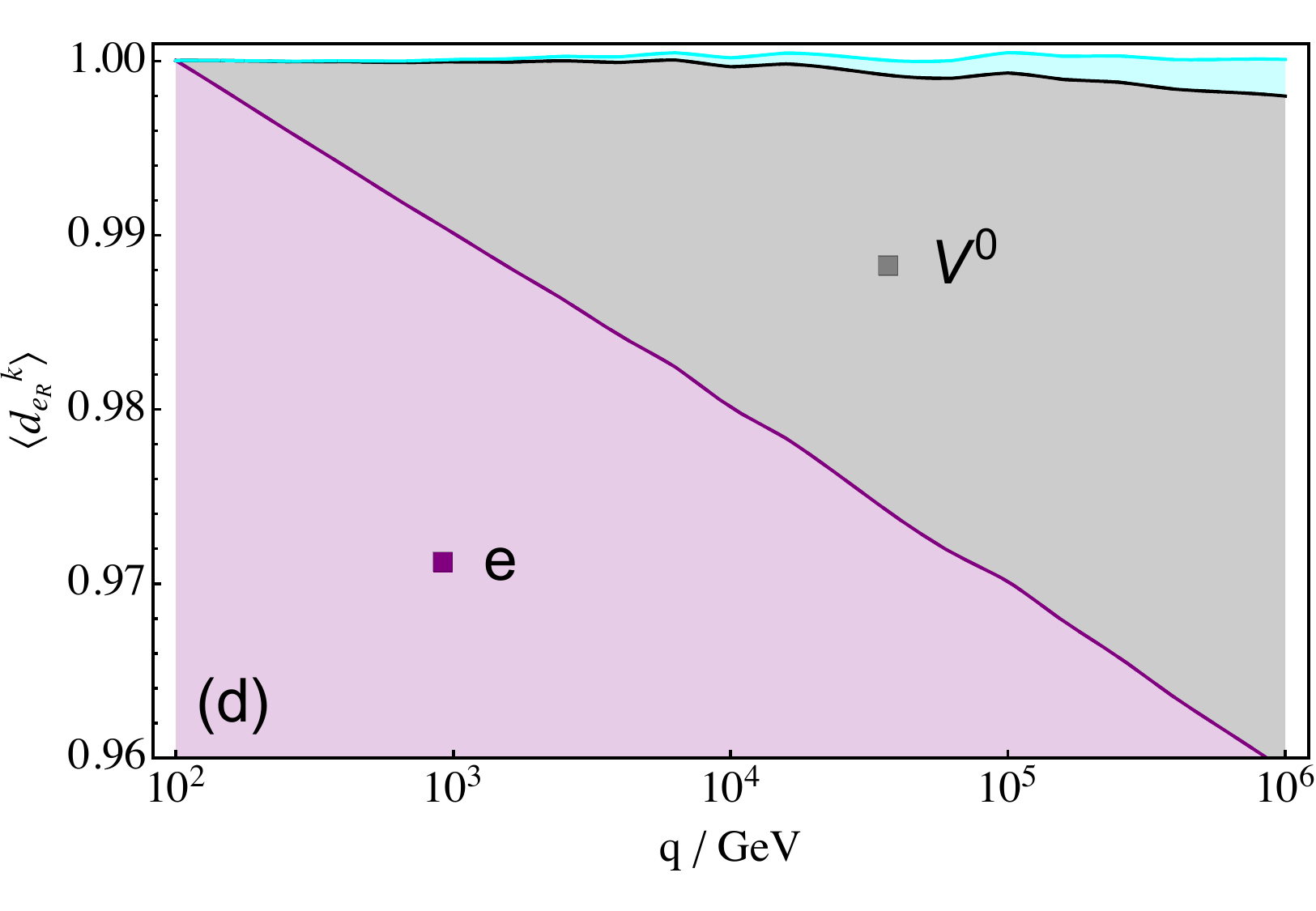}
  \includegraphics[scale=0.42]{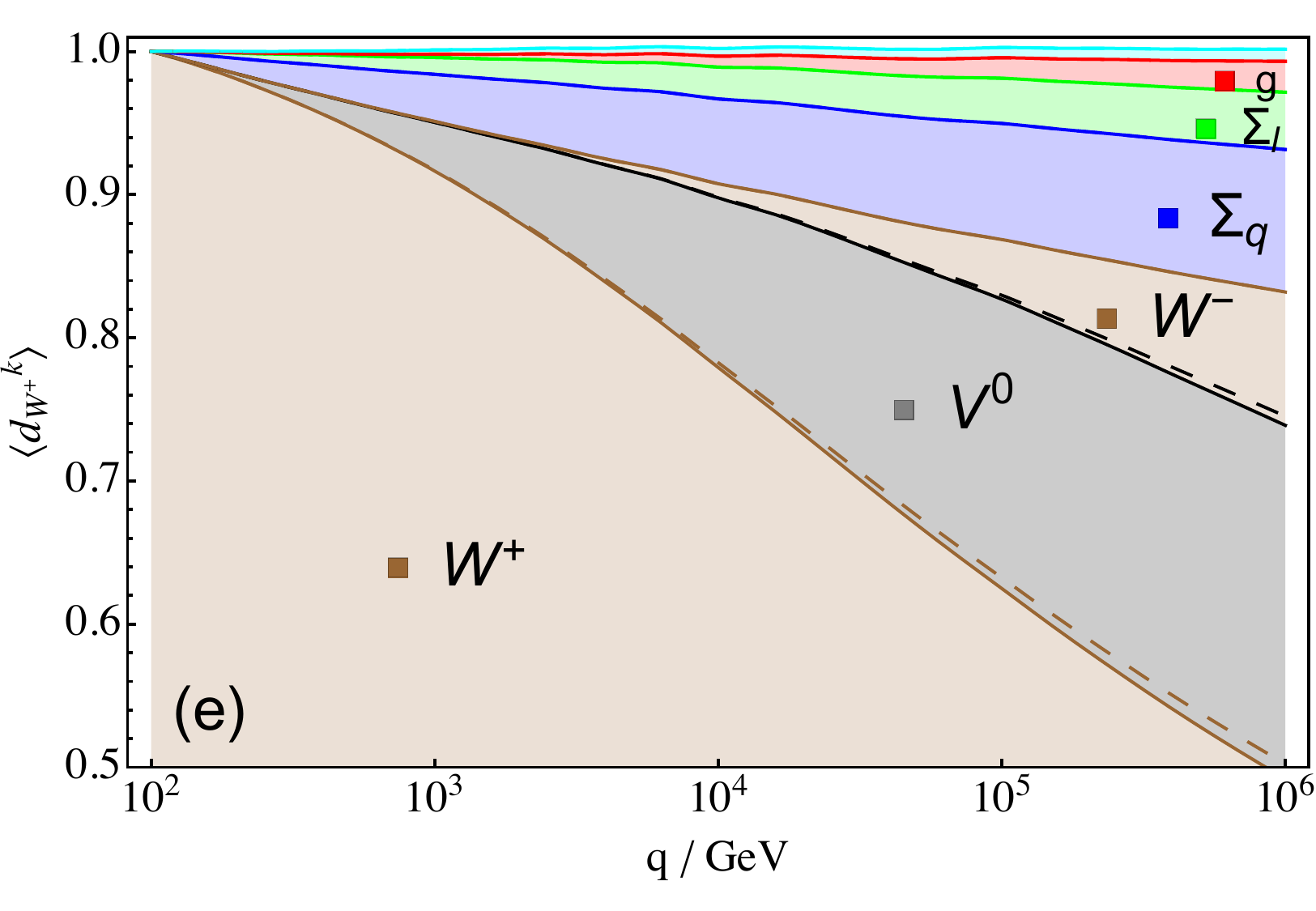}
  \includegraphics[scale=0.42]{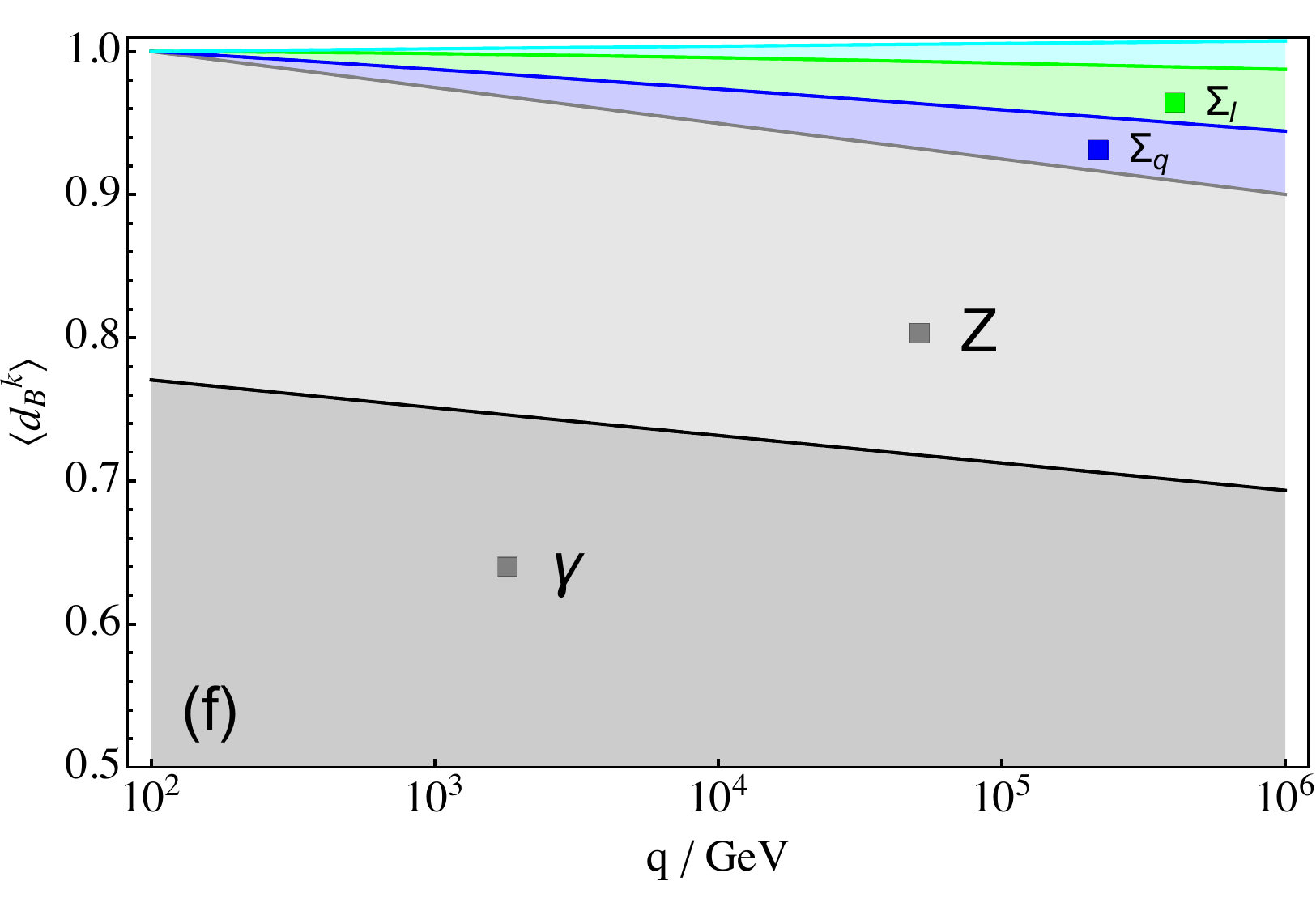}
 \includegraphics[scale=0.42]{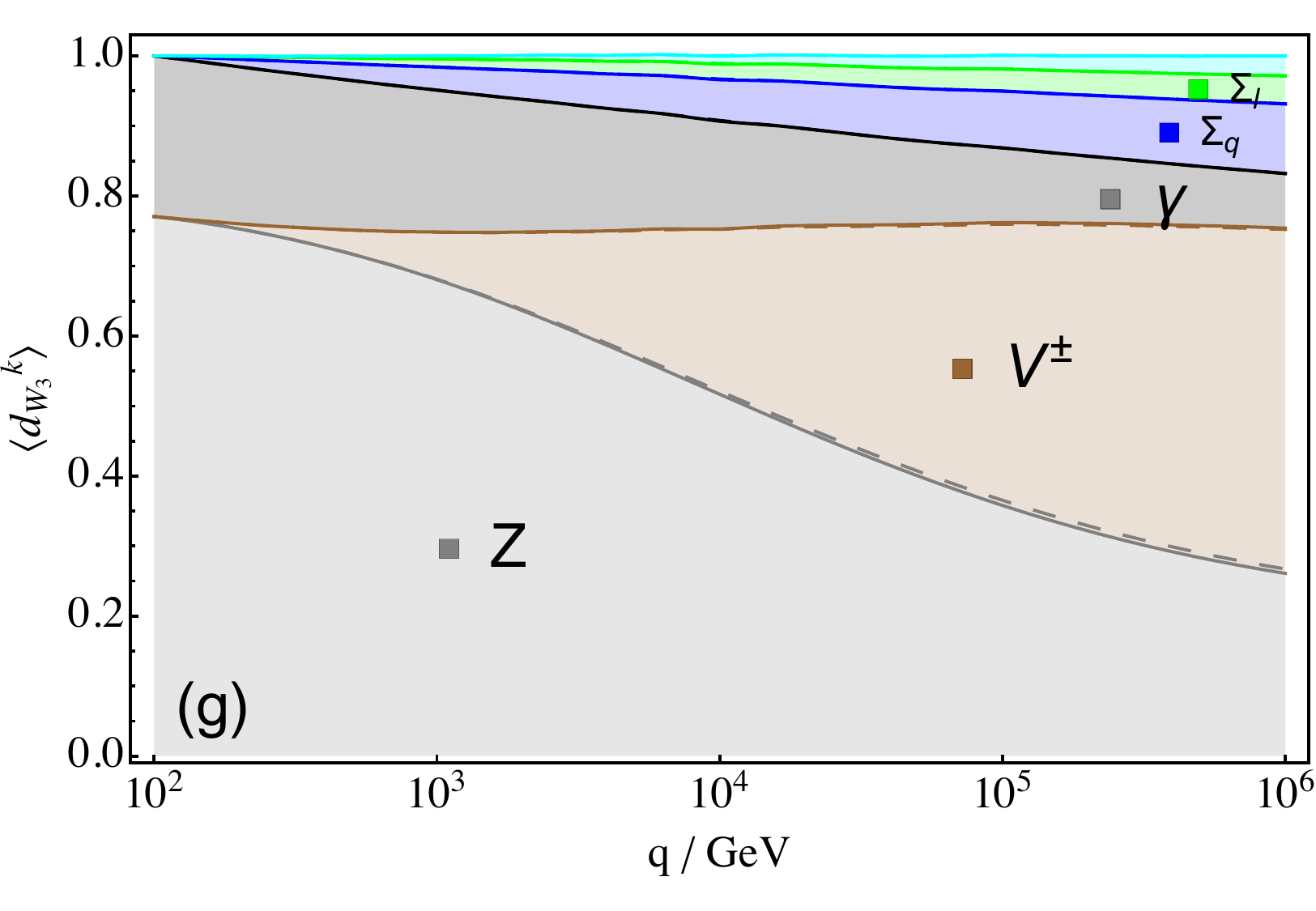}
 \includegraphics[scale=0.42]{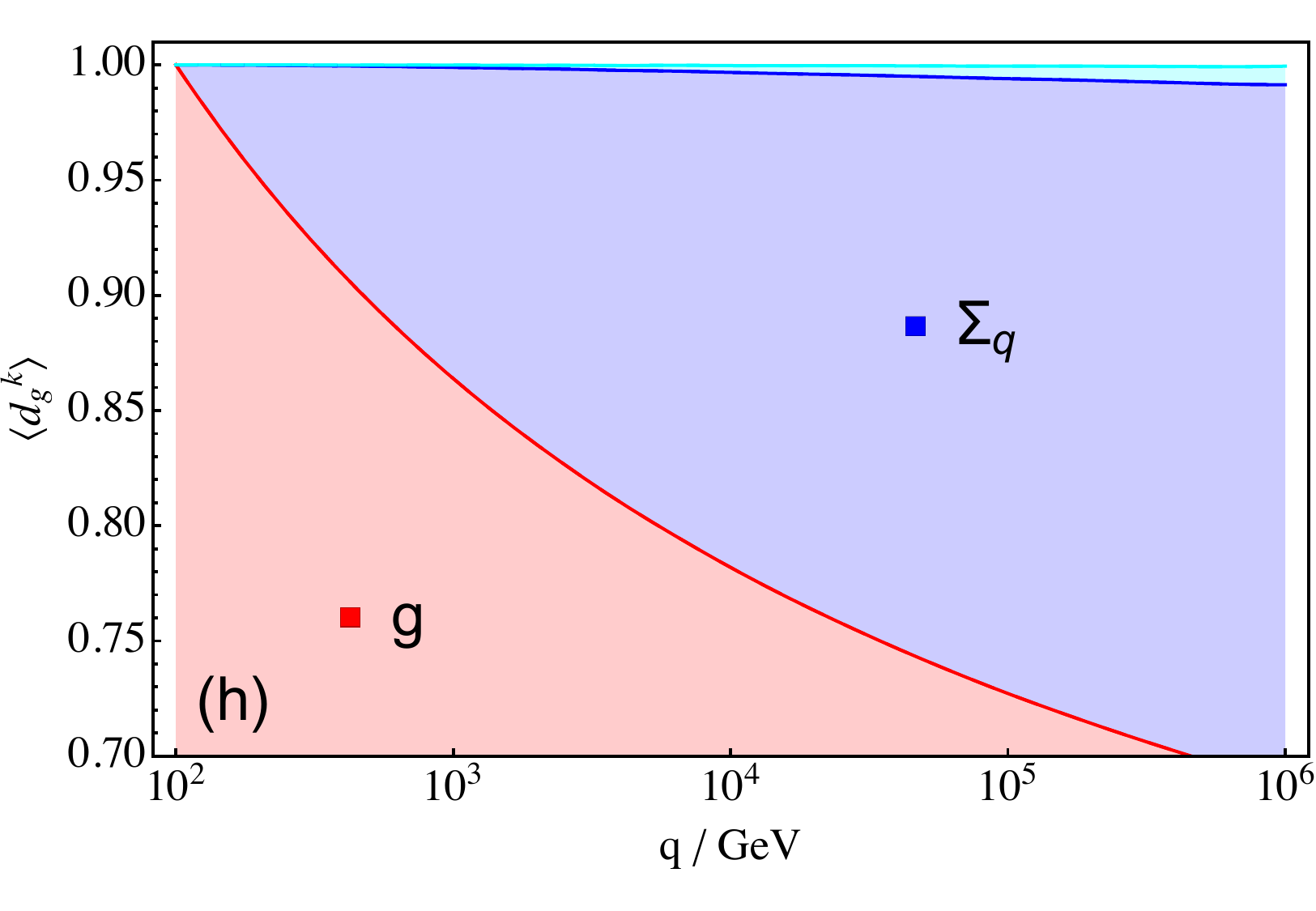}
\caption{\label{fig:Fig1}%
The momentum averaged fragmentation functions $\vev{d^k_i}$ for (a,b)
$i = d_L, d_R$, (c,d) $e_L, e_R$, (e,f) $W^+,B$,
(g,h) $W_3, g$. The different values of $k$ are stacked on top of
each other, such that the total equals one, as demanded by the sum
rule. Dashed/solid lines show DL/NLL resummed results. }
}
\FIGURE[h]{
 \centering
  \includegraphics[scale=0.43]{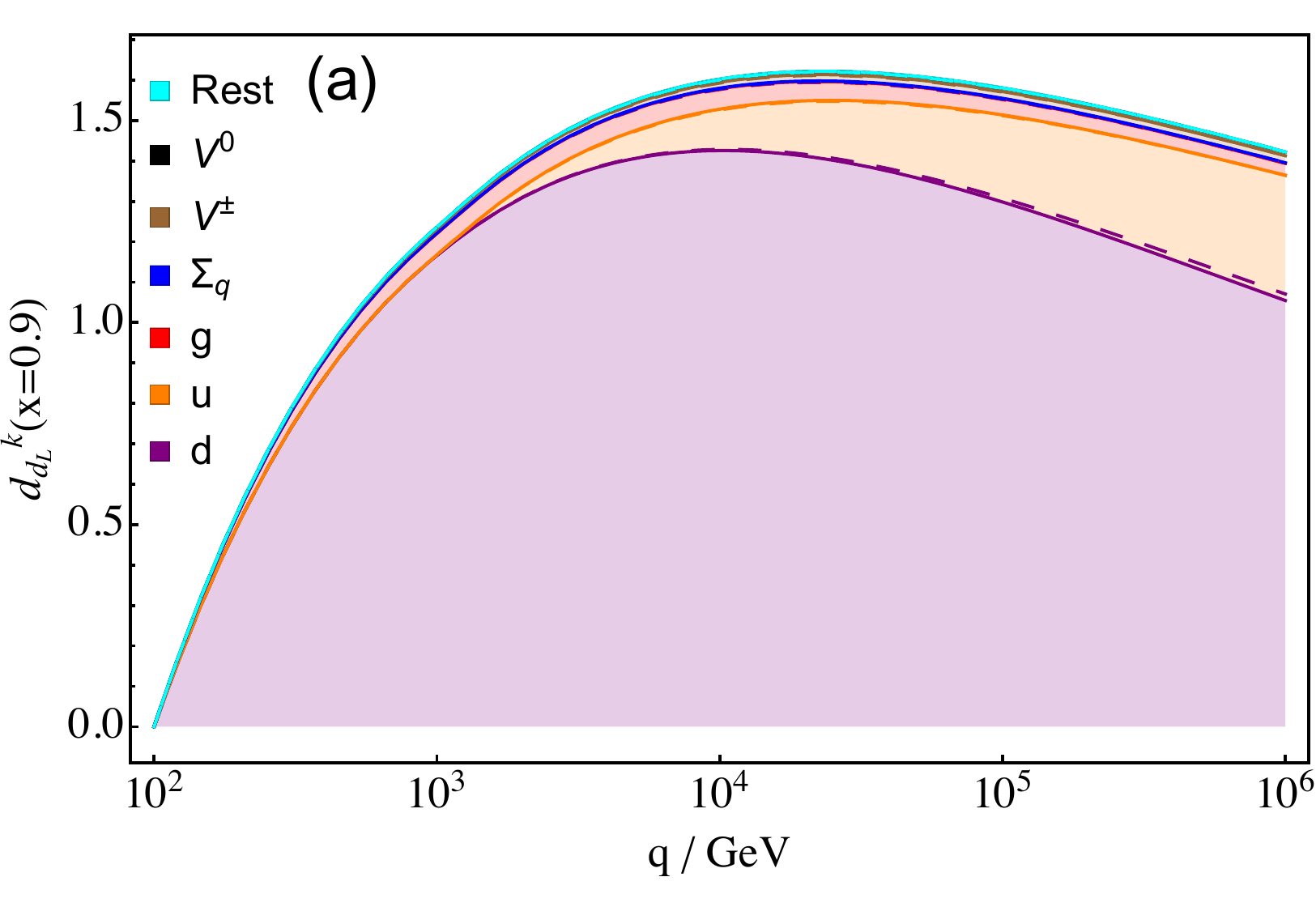}
  \includegraphics[scale=0.43]{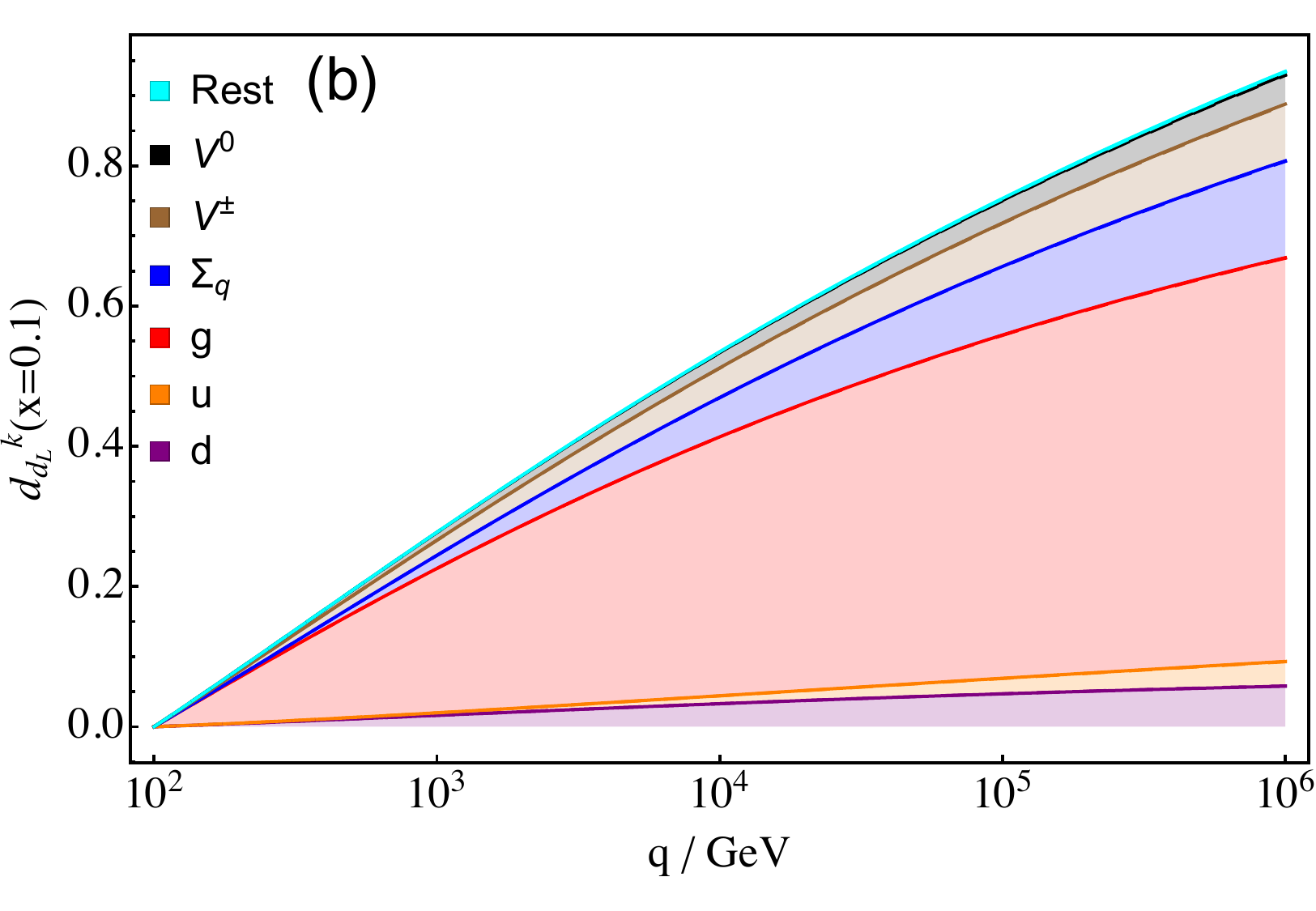}
  \includegraphics[scale=0.43]{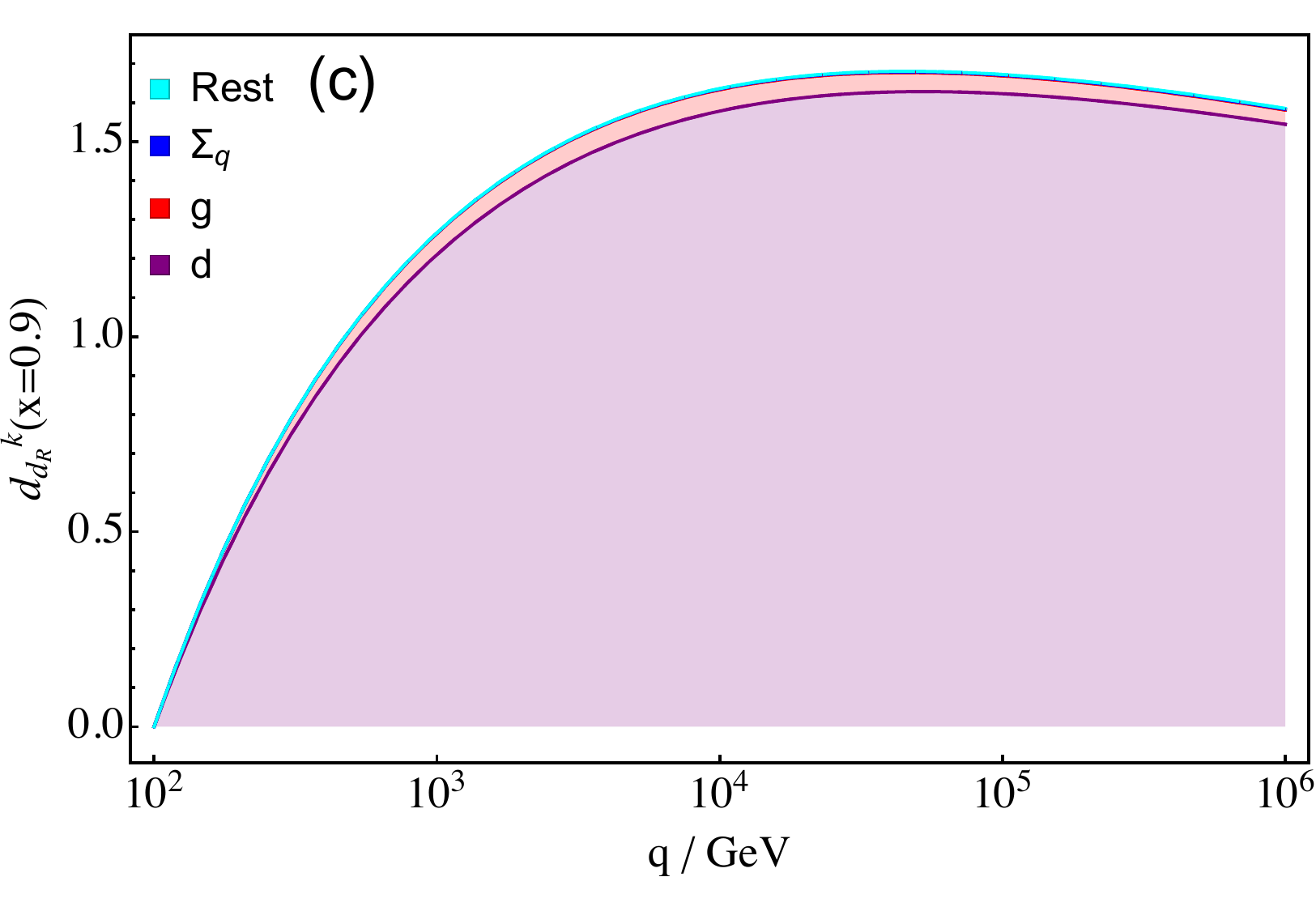}
  \includegraphics[scale=0.43]{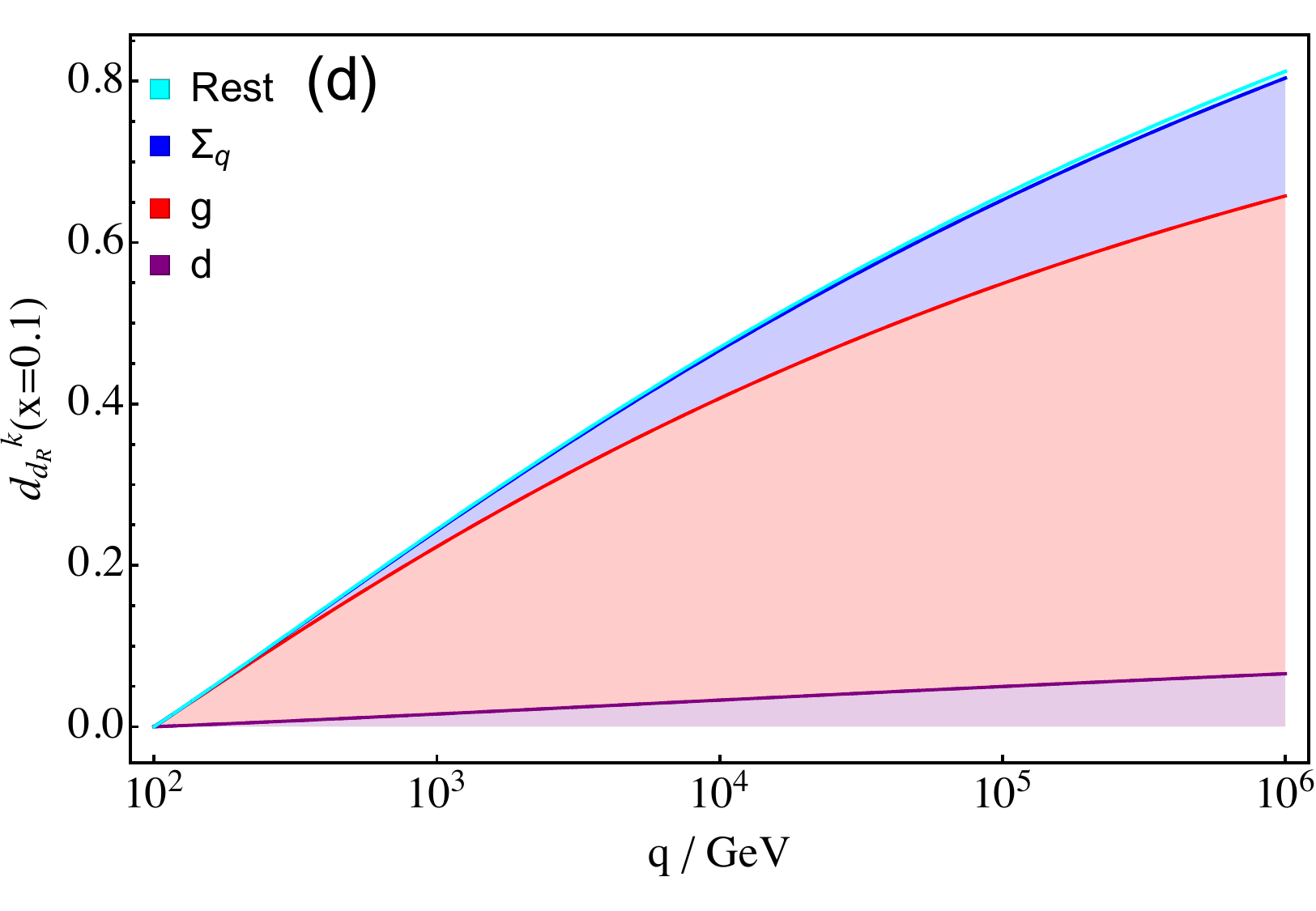}
  \includegraphics[scale=0.43]{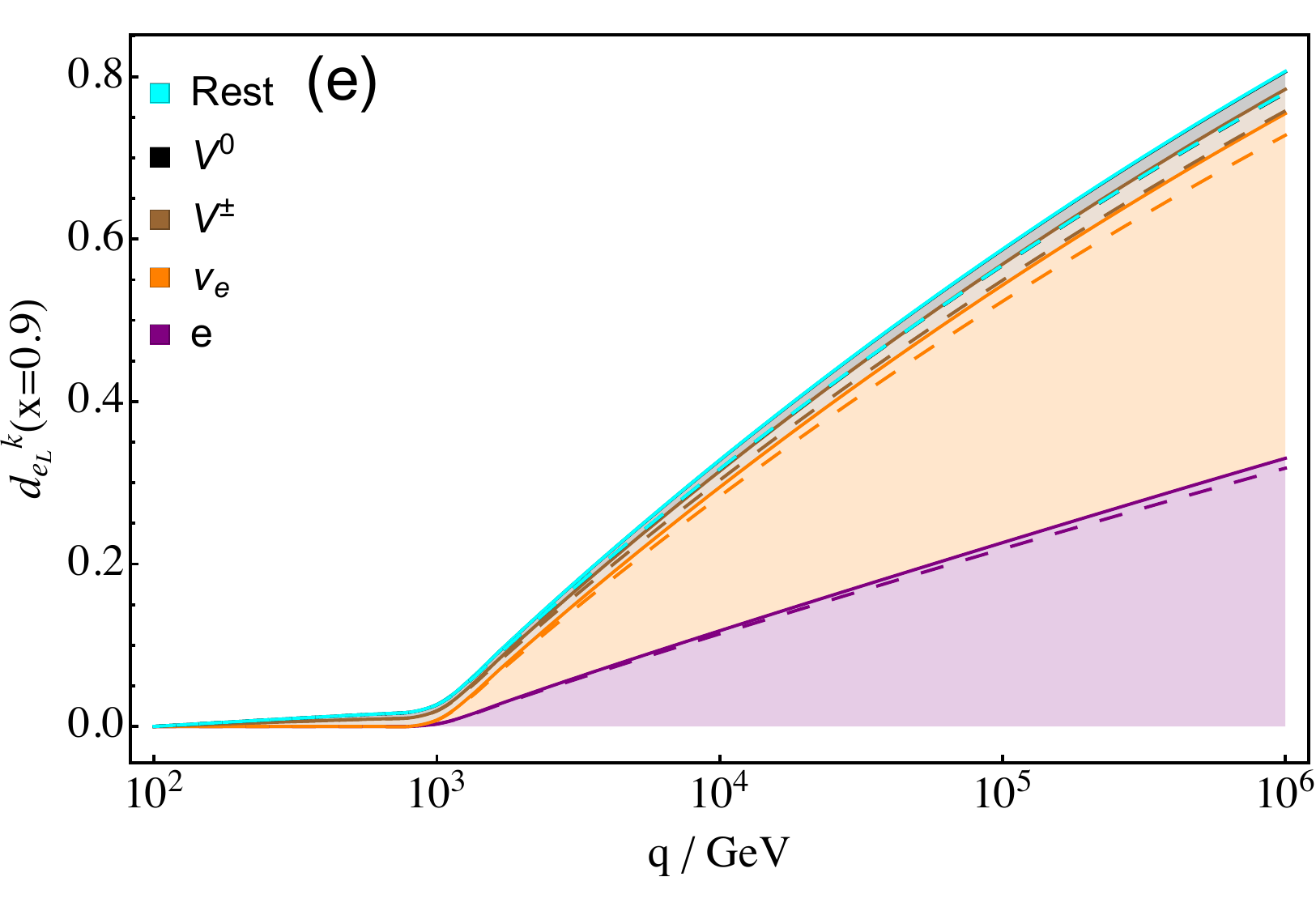}
  \includegraphics[scale=0.43]{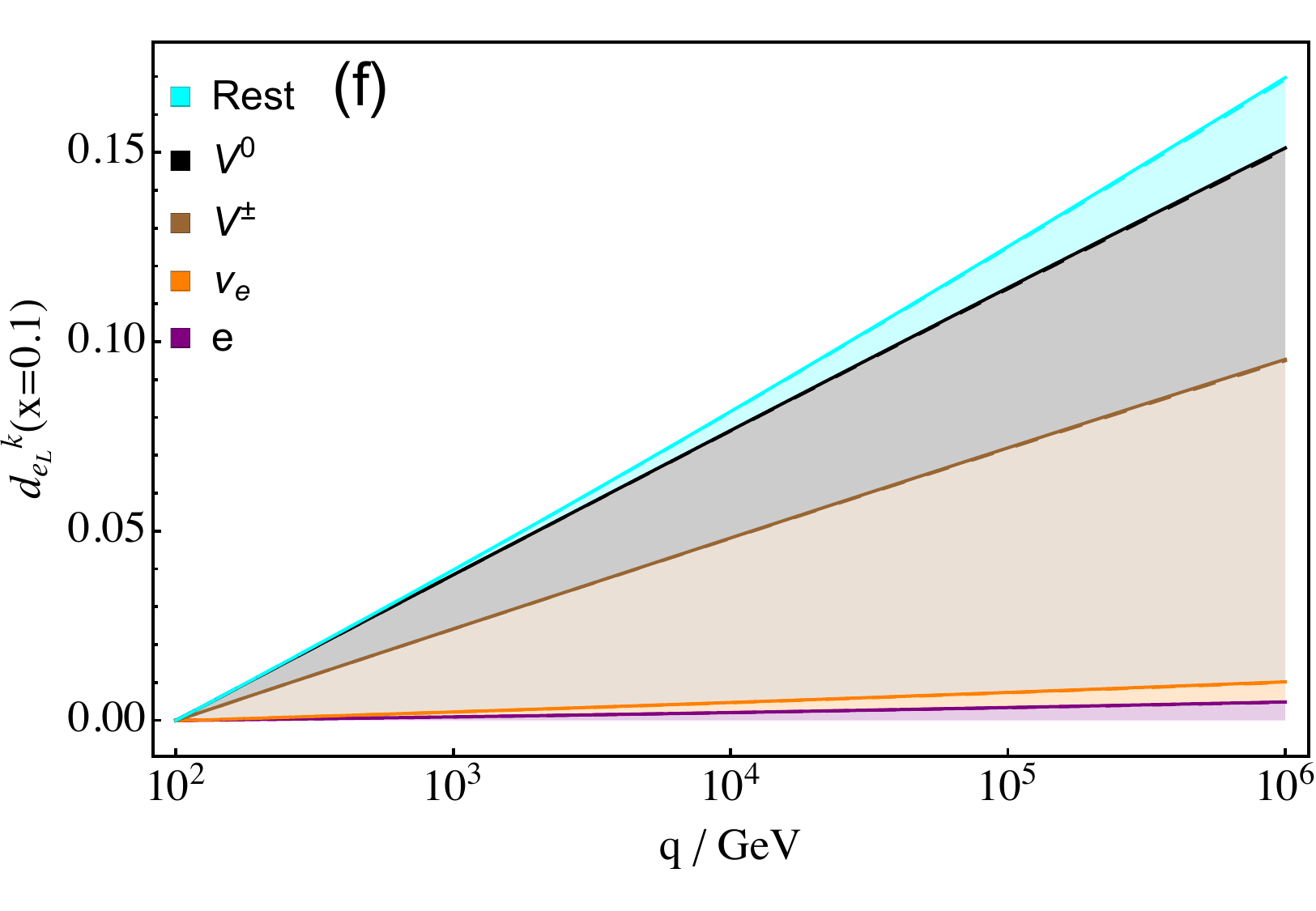}
  \includegraphics[scale=0.43]{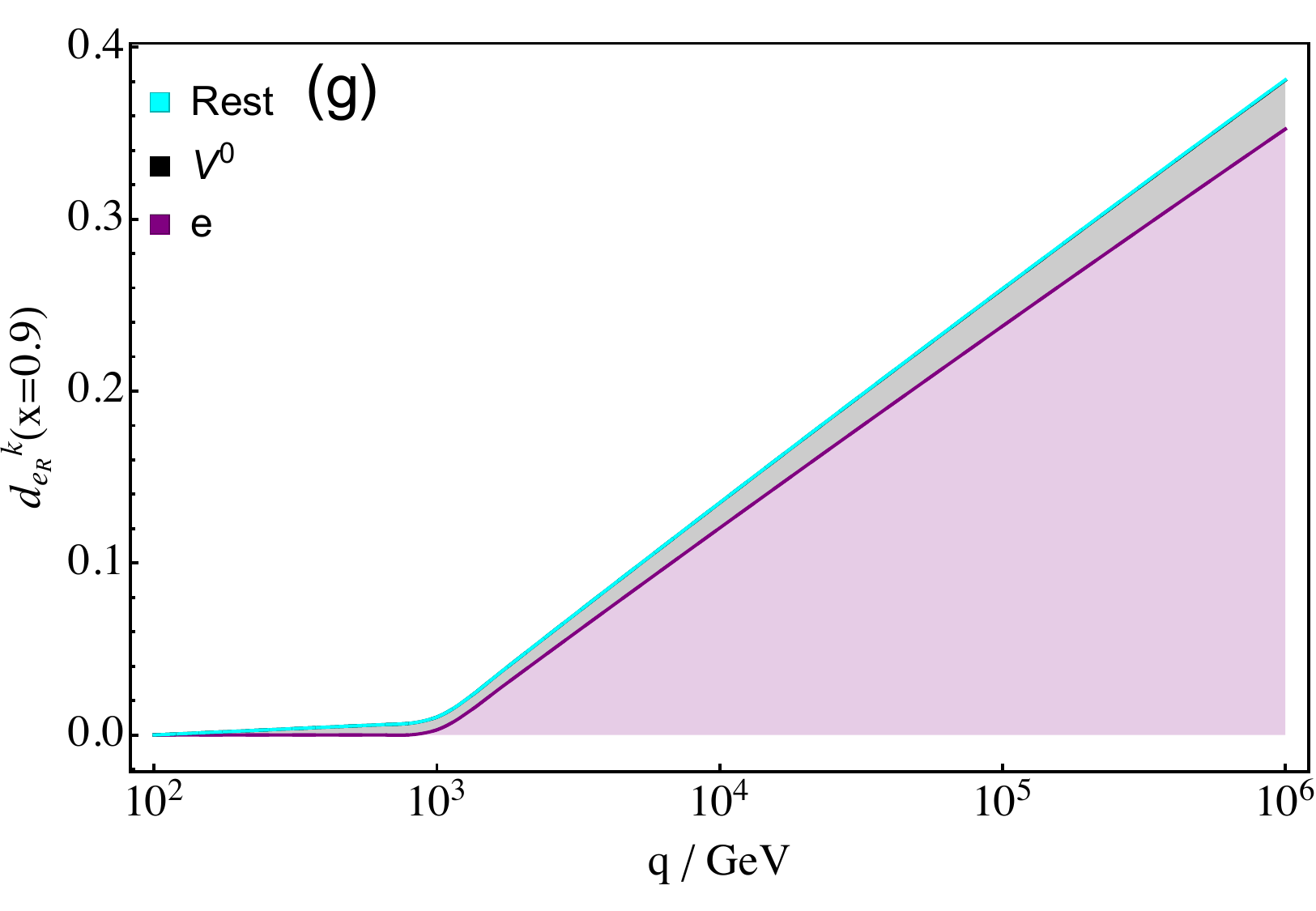}
  \includegraphics[scale=0.43]{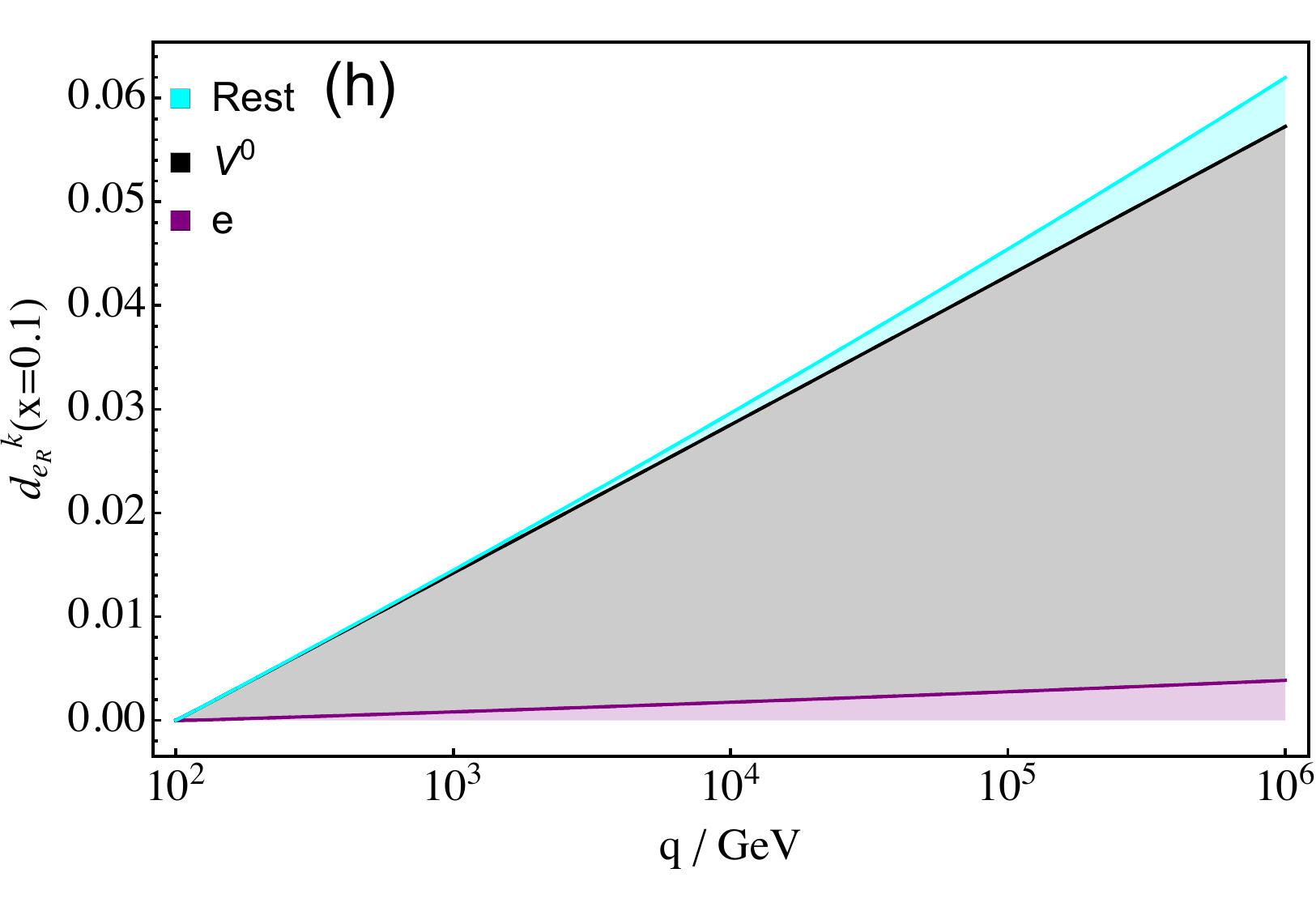}
\caption{\label{fig:Fig2}%
The fragmentation functions at $x = 0.9$ and $x = 0.1$ for (a,b) $i = d_L$,
(c,d) $d_R$, (e,f) $e_L$, (g,h) $e_R$. The different values of $k$ are stacked on top of each
other. Dashed/solid lines show DL/NLL resummed results. }}
\FIGURE[h]{
 \centering
  \includegraphics[scale=0.43]{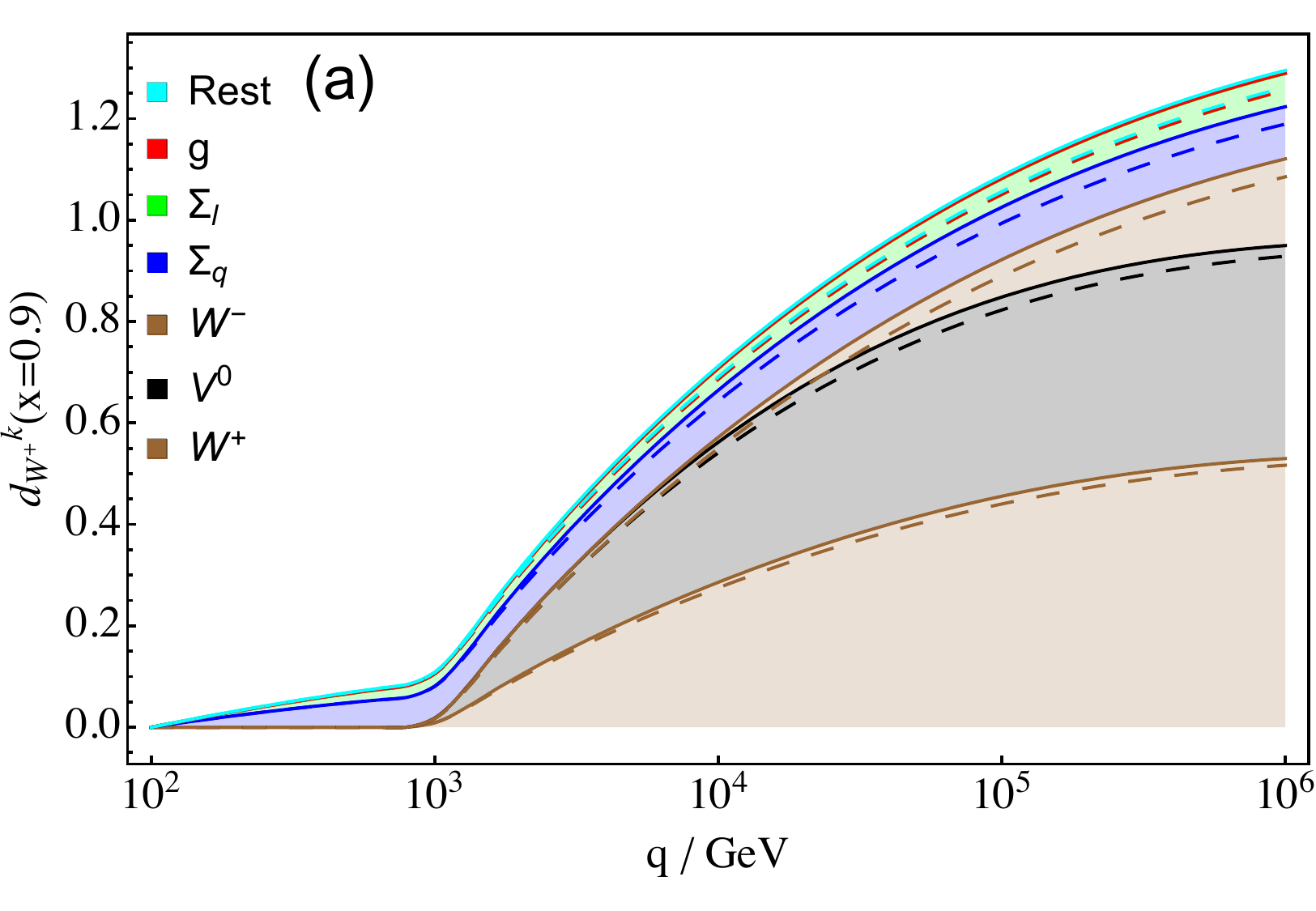}
  \includegraphics[scale=0.43]{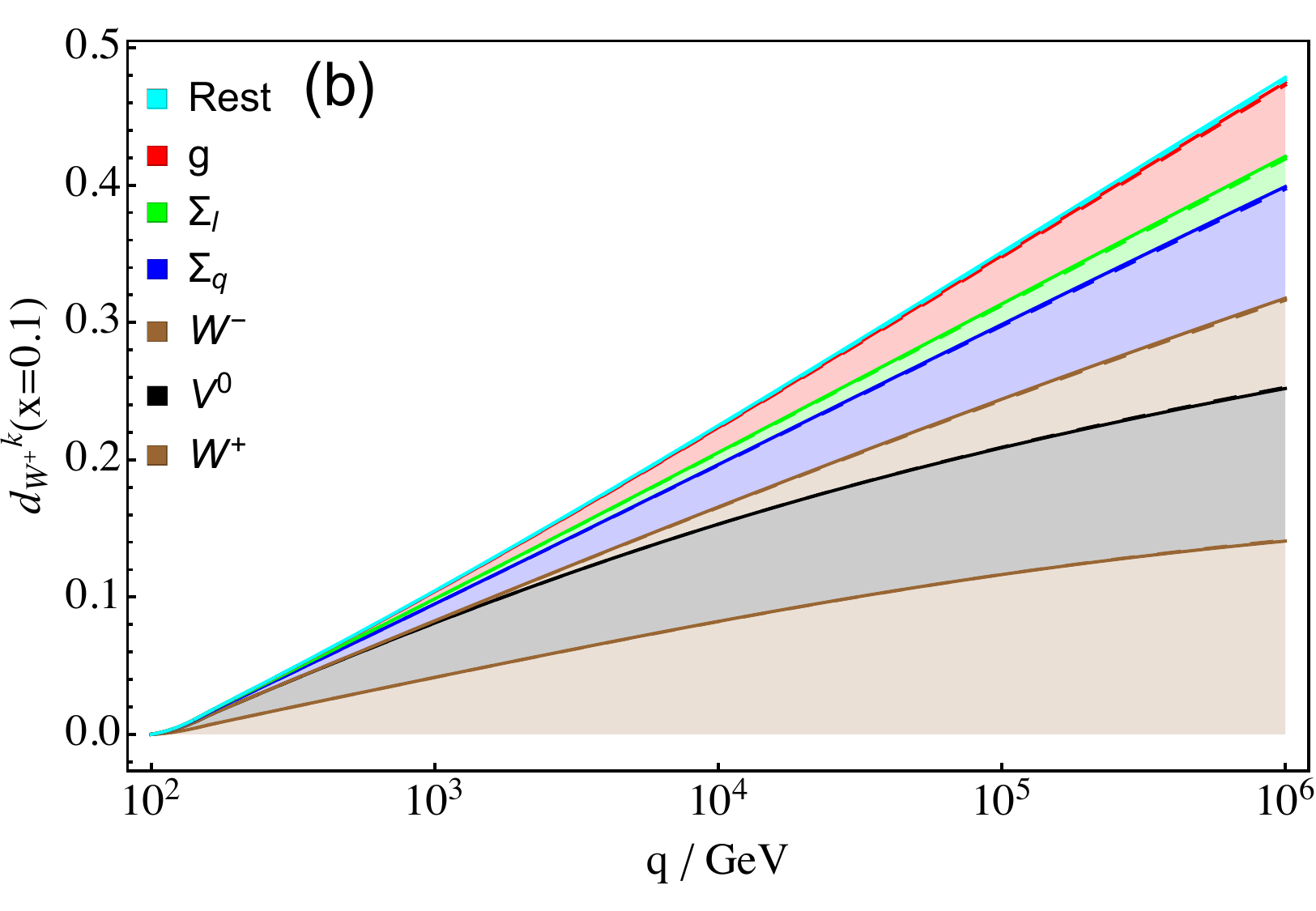}
  \includegraphics[scale=0.43]{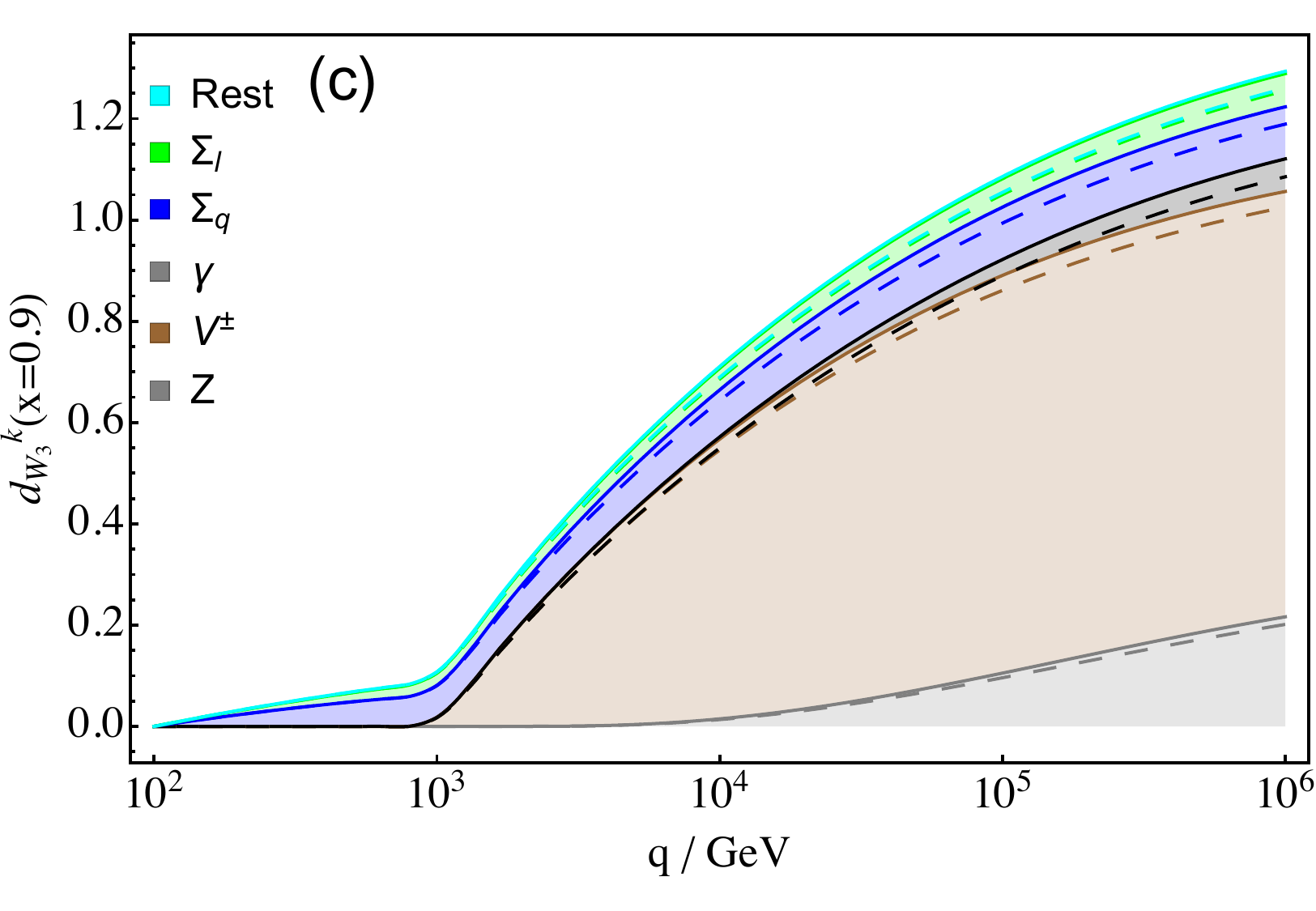}
  \includegraphics[scale=0.43]{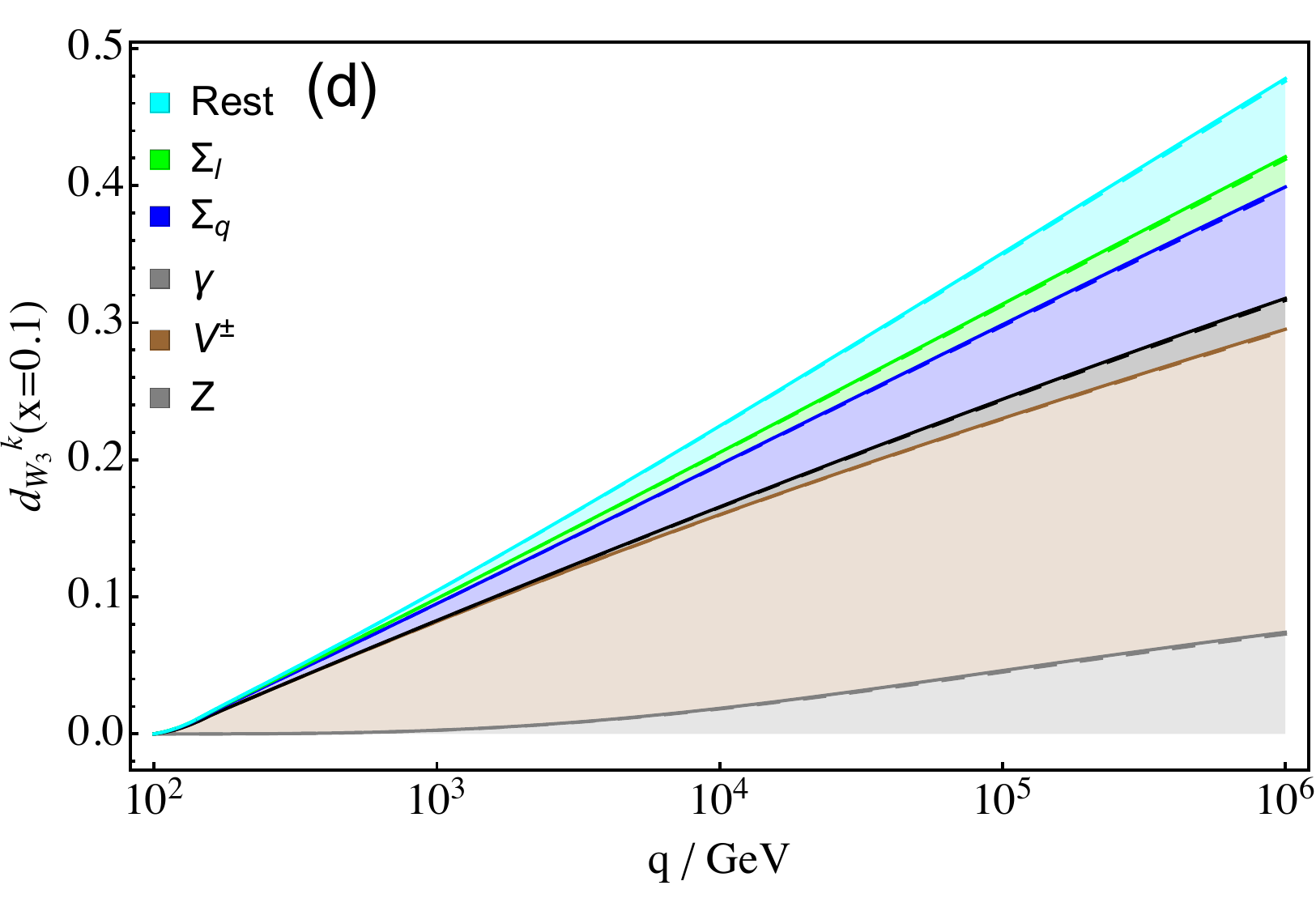}
  \includegraphics[scale=0.43]{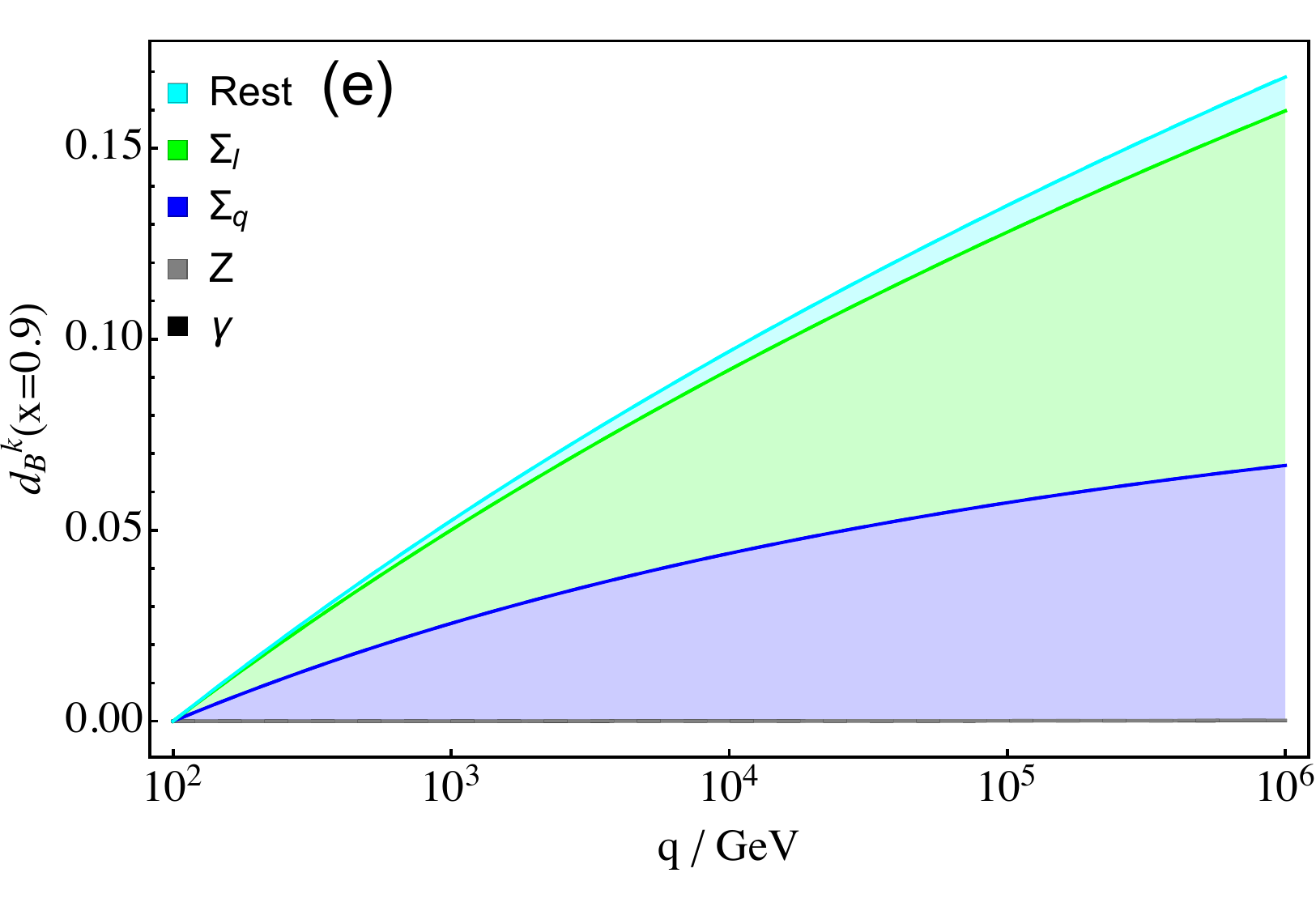}
  \includegraphics[scale=0.43]{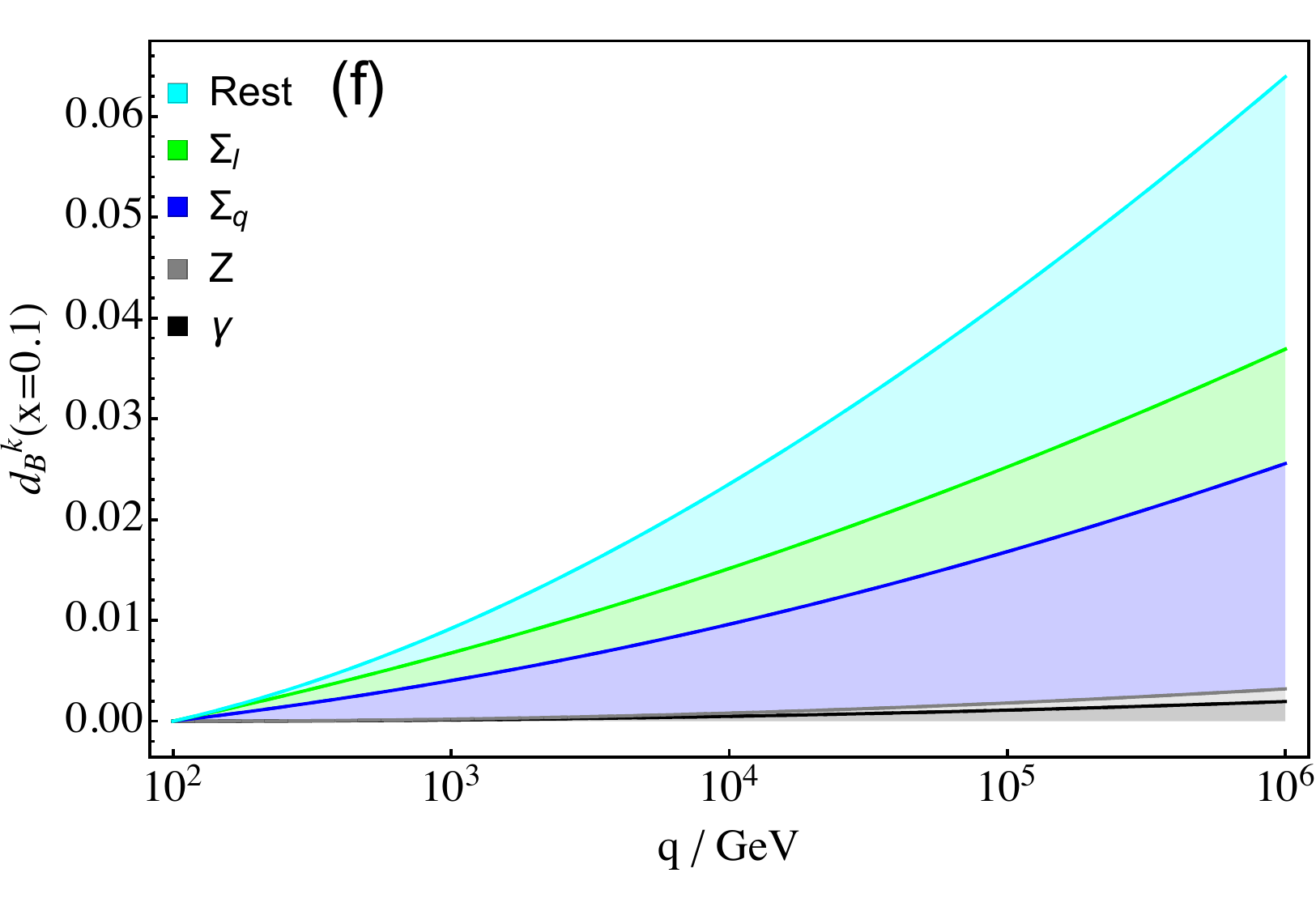}
  \includegraphics[scale=0.43]{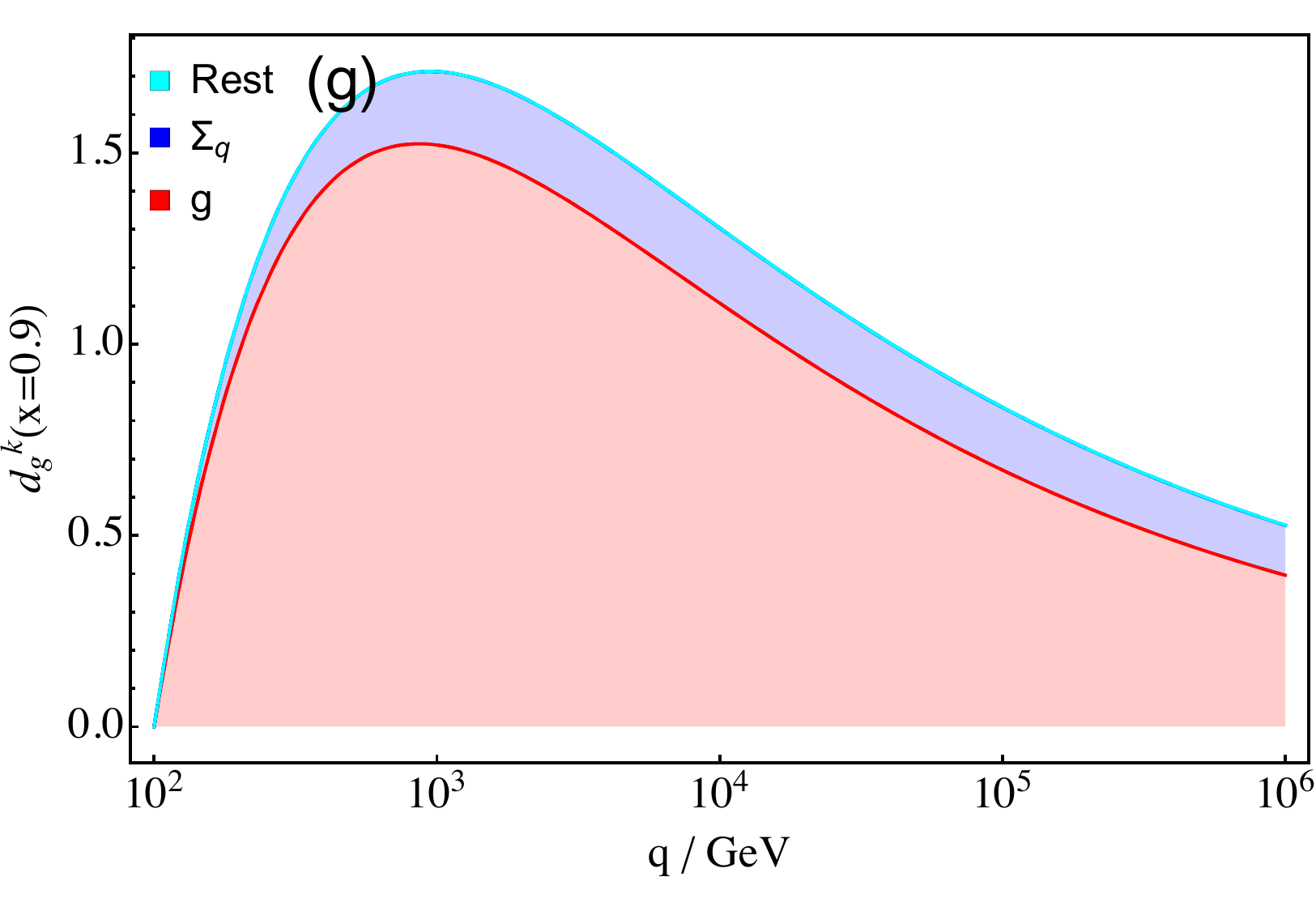}
  \includegraphics[scale=0.43]{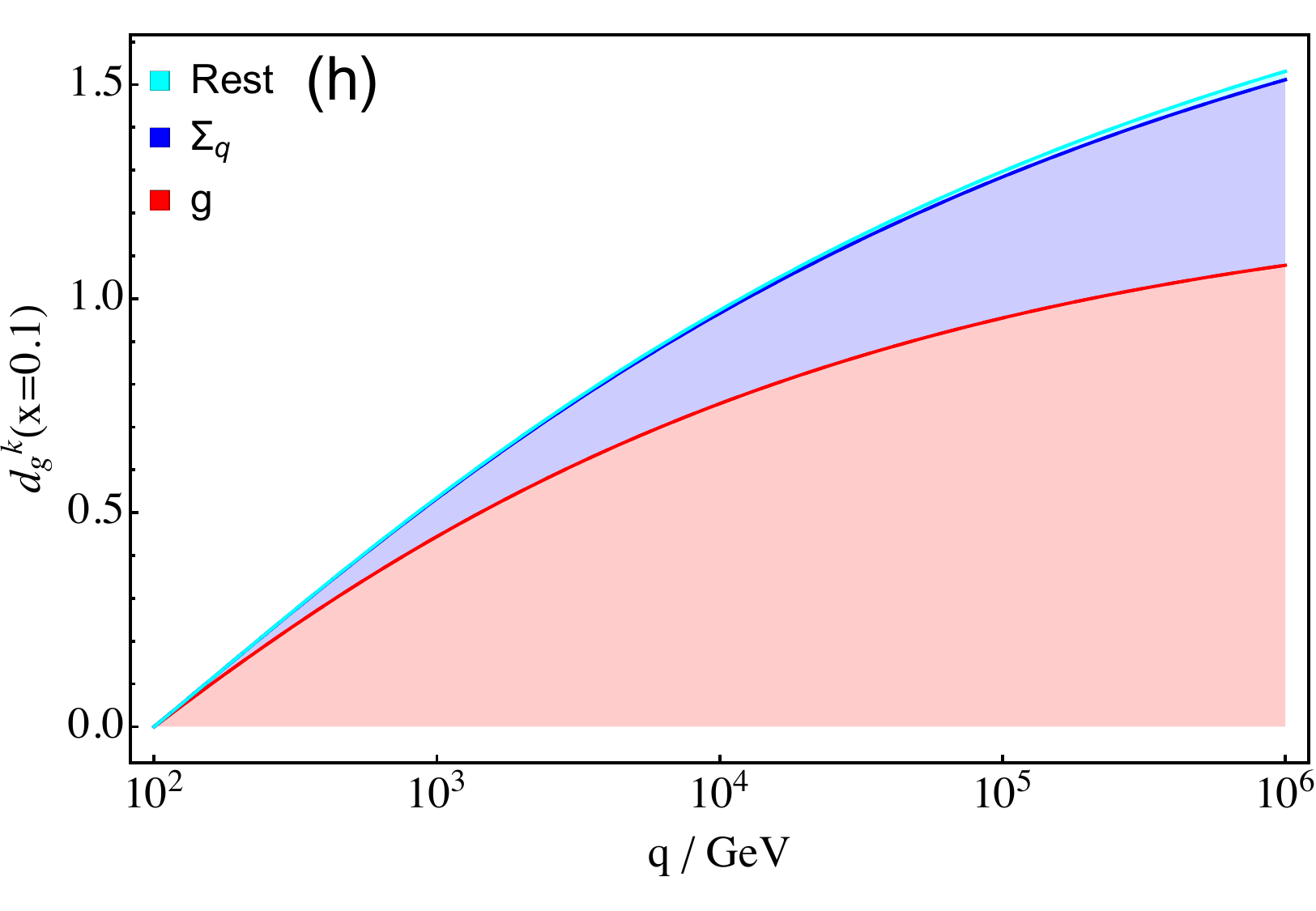}
\caption{\label{fig:Fig3}%
The fragmentation functions for $x = 0.9$ and $x = 0.1$ for (a,b) $i = W^+$,
(c,d) $W^3$, (e,f) $B$, (g,h) $g$. The different values of $k$ are stacked on top of each
other. Dashed/solid lines show DL/NLL resummed results. }}
\FIGURE[h]{
 \centering
  \includegraphics[scale=0.43]{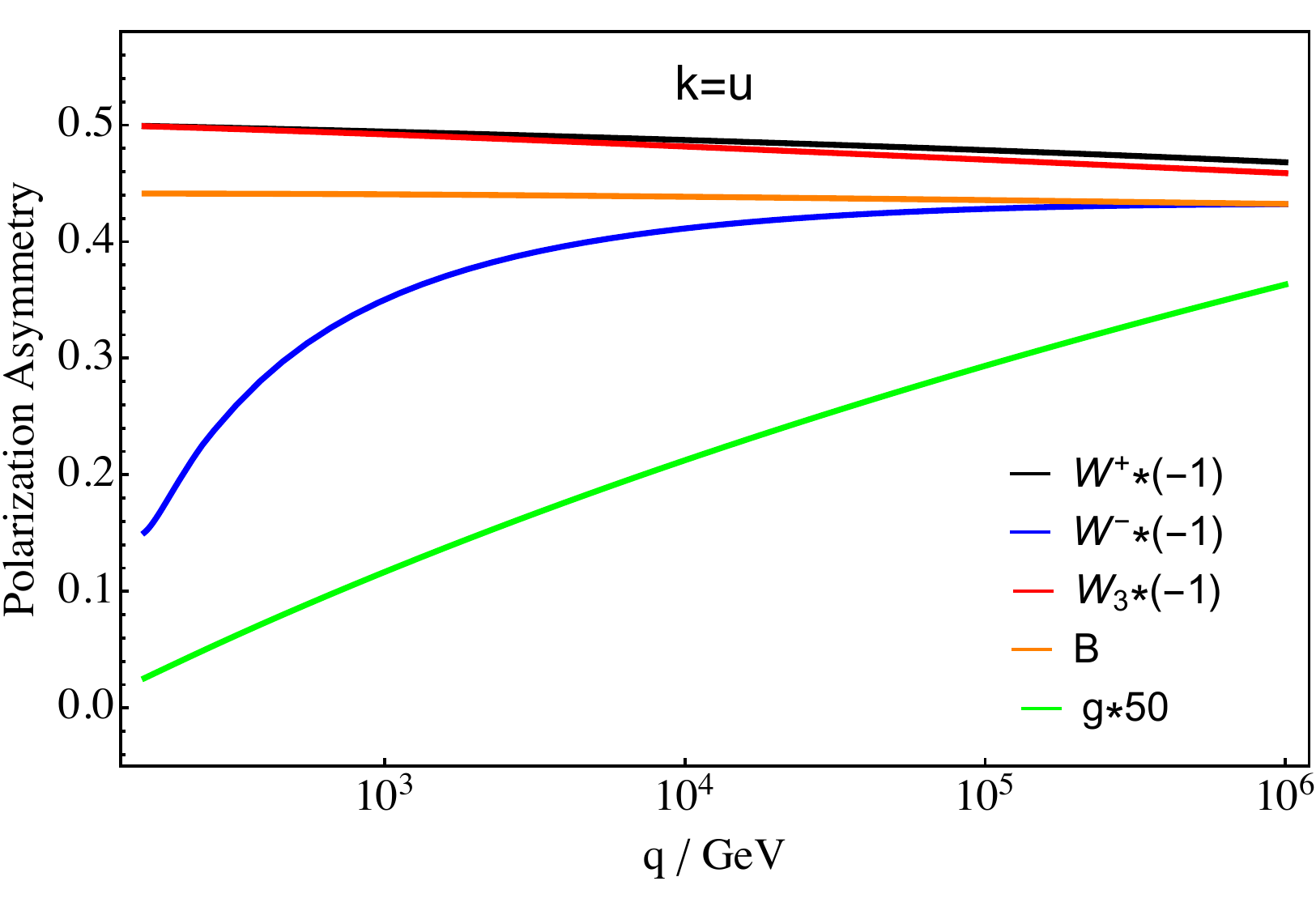}
  \includegraphics[scale=0.43]{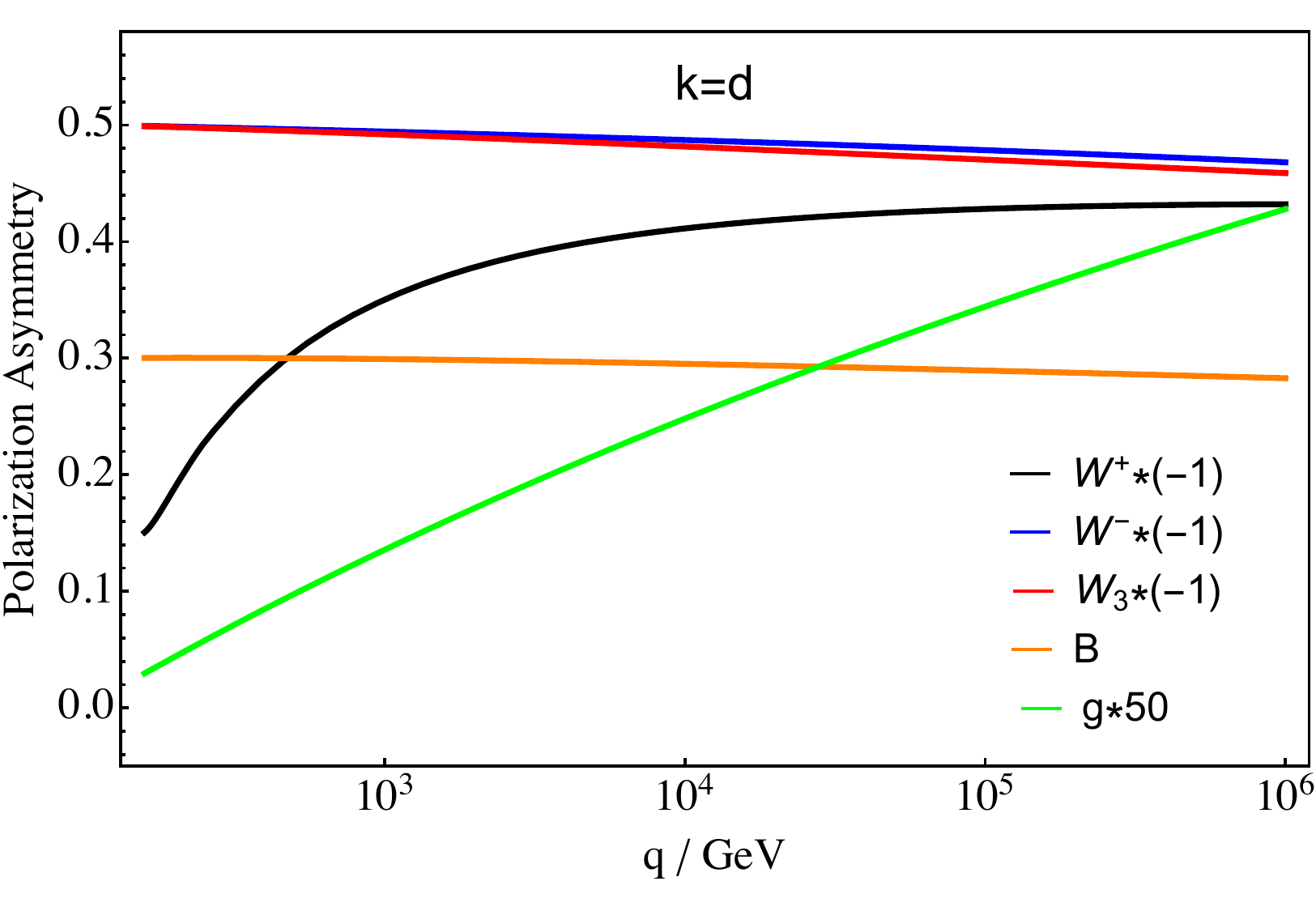}
 \caption{\label{fig:Fig4}%
The absolute value of the polarization asymmetry, defined as $A(V) =
(\vev{d_{V_+}^k} - \vev{d_{V_-}^k}) /  (\vev{d_{V_+}^k} +
\vev{d_{V_-}^k})$, for fragmentation into 
(left) $u$ and (right) $d$ quarks, for the vector
bosons $W^\pm$, $W_3$, $B$ and the gluon.
Note that the gluon asymmetry is scaled
by a factor of 50, and that for the SU(2) bosons the negative of the asymmetry is shown. The results use the NLL accuracy as discussed in Section~\ref{sec:leading_log_2}.}}
As already mentioned, there are a total of $58\times30 = 1740$
distinct FFs, and we can clearly only show a small subset of all
possible results. We therefore choose a few illustrative choices of
$i$ (left-and right-handed down quarks, the left- and right-handed
electron, the SU(2) bosons $W^+$ and $W_3$, the U(1) boson $B$ and the
gluon), and for each $i$ group the 30 possible values of $k$ into a
few representative sets. Readers interested in more details can
request all data as LHAPDF files from the authors. The main results
use the full NLL accuracy of the DGLAP evolution that
was discussed in section~\ref{sec:leading_log_2}. Note that to obtain
full NLL accuracy for a cross section prediction requires the
inclusion of single logarithms arising from the evolution of the soft
function that were computed in~\cite{Manohar:2018kfx}.  

We begin by showing in Fig.~\ref{fig:Fig1} the results for the
momentum fractions $\vev{d_i^k(q)}$ defined in \eq{momFracDef}. For
each species $i$, we show how the total momentum is shared between
fragmentation particles
$k$ at scales $q$ ranging from 100 to $10^6$ GeV. We stack the various
sets for $k$ on top of each other, such that momentum conservation
implies that each plot sums to unity for all values of $q$ once all
particles are included. To show the size of the difference between DL
and NLL evolution, we show in dashed lines also the results obtained
using DL evolution. The reason that for several curves the DL result
is not visible is because it is indistinguishable from the NLL
result. One can also clearly see that at $q = 100$ GeV, the only
contribution is for $i = k$. Since $i$ and $k$ are chosen in the
unbroken and broken basis, respectively, for the $W_3$ and $B$ the
relative probability of $Z^0$ and $\gamma$ are given by the weak
mixing angle. As we evolve to larger values of $q$, other flavors $k$
are generated.

In the first row we show the fragmentation of left- and right-handed down
quarks, $i=d_L,d_R$. In the left-handed case (a) one can see that the
SU(2) interaction has a significant effect.  Left-handed up quarks are
generated with double logarithmic probability, such that at large
enough values of $q$ the amount of $u_L$ and $d_L$ become of the same
order of magnitude, and SU(2) bosons are produced at an appreciable
rate. Gluons are produced at a larger rate, which is
obviously due to the relative strength of the SU(3) and SU(2)
interactions. For the right handed down quark (b), the fragmentation is
completely dominated by QCD evolution, such that a large fraction
of gluons and a smaller fraction of quarks other than $d_R$ get
generated. Other particles, shown by the remaining contribution in
cyan, only make up a tiny fraction, even at $q = 10^6$ GeV.

The fragmentation of left- and right-handed electrons is shown in the
second row of Fig.~\ref{fig:Fig1}. In the left-handed case (c) one
can again see the importance of the SU(2) interactions at large
values of $q$, and for $q \sim 10^6$ GeV the relative fraction of
electrons and neutrinos becomes comparable, with the momentum fraction
contained in gauge bosons at the 10\% level.  For the right-handed
electron (d), the evolution is only given by the U(1)
interaction, such that one generates only a small fraction of U(1)
bosons, and an even smaller fraction of other particles. 

Gauge boson fragmentation is shown in the third and fourth rows of
Fig.~\ref{fig:Fig1}. 
For the $W^+$ boson (e), one sees that the other SU(2) gauge bosons
are generated quite rapidly as $q$ becomes larger than 100
GeV. Asymptotically, for $q \to \infty$, the three SU(2) gauge bosons
will evolve to have equal momentum fractions, and while one can see
the trend for them to become equal, one needs to go to much higher
values than are shown here. Quarks and leptons are also produced at an
appreciable rate, with more quarks owing to the colour factor.
For the U(1) boson (f), the only non-vanishing fragmentation at $q=
100$ GeV is into $Z$ bosons and photons, with relative fraction
$\tan^{-2}\theta_W$. Since the coupling constant $\alpha_1$ is
smaller than $\alpha_2$, quarks and leptons are produced at a lower
rate than for the $W^+$ boson.  However, the quark and lepton rates are
more equal, because the colour factor of the quarks is largely
compensated by the higher hypercharges of the leptons. 
For an initial $W_3$ boson, shown in (g), one again starts off with
only $Z$ bosons and photons, with relative fraction
$\tan^2\theta_W$. Quickly the neutral SU(2) boson evolves into charged
$W$s, and also into quarks and leptons. Finally, we show in (h) the
evolution of the gluon. As expected, it is completely dominated by the
strong interaction, such that it mostly evolves into quarks.

While Fig.~\ref{fig:Fig1} illustrates the evolution of the total
momentum fractions carried by various particles in a given species of jet, it does not show how
the evolution looks for fixed values of $x$. This is shown in
Figs.~\ref{fig:Fig2} and~\ref{fig:Fig3} for the same set of particles
as before, and using the values $x = 0.9$ (shown on the left) and $x =
0.1$ (on the right). As in Fig.~\ref{fig:Fig1}, solid (dashed) lines correspond to NLL (DL) evolution. As explained in
Section~\ref{sec:implementation}, the initial condition at $q = 100$
GeV is a $\delta$-function at $x = 1$ for $i = k$, such that the
fragmentation at $x = 0.9$ is overall much more dominated by $k = i$
than at lower values of $x$.  Notice that the constraint $x<1-m_V/q$
for emission of a heavy vector boson means that at $x=0.9$ no such
emission can occur below $q=m_V/(1-x)\sim 1$ TeV, depressing
the evolution of leptons and heavy bosons below that scale.  At 
 $x = 0.1$, fragmentation into vector bosons is dominant at all
 scales, because of the low-$z$ enhancement of the corresponding
splitting functions.

In  Fig.~\ref{fig:Fig4} we show some results on the polarization
asymmetry of vector bosons fragmenting into up and down quarks.
For $W^+,W_3\to u$ and $W^-,W_3\to d$ the asymmetry is large and
negative, due to the dominance of the $V_-\to f_L$ splitting function.
The $W^-\to u$ and $W^+\to d$ asymmetries are also negative but
increase from zero as they start at higher order.  The $B$ asymmetries
are positive because of the dominance of fragmentation into the
right-handed quarks in that case.  The gluon asymmetry is a secondary
effect of the different evolution of left- and right-handed quarks,
the latter evolving more slowly and so remaining at higher $x$.
Notice that there is even a slight difference between the gluon
asymmetries for fragmentation into up and down quarks, due to their
different electroweak evolution.

Although substantial vector bosons polarizations are generated by
electroweak evolution, their effects on fragmentation into fermions
and unpolarized bosons are negligible.  The boson helicity asymmetries
$d^k_{V_+} - d^k_{V_-}$ start from zero at the electroweak scale and 
cannot affect the unpolarized bosons at all, as they have opposite CP
quantum numbers. They can indirectly affect only the
$\{\mathbf T, \mathrm{CP}\}=\{0,-\}$ and $\{1,+\}$ fermion FFs, generally
producing effects at the $10^{-4}$ level or less in the momentum-averaged
FFs of individual fermions.

\section{Conclusions}
\label{sec:Conclusions}
In this paper we have discussed the evolution of fragmentation
functions in the full Standard Model, which requires resummation of
leading logarithms arising from final-state radiation and the
associated virtual corrections.

At energy scales far above the electroweak symmetry breaking scale,
short distance processes can be described in terms of  58 particles
in the unbroken Standard Model: 12 left-handed quarks, 12 right-handed
quarks, 12 left-handed leptons, 6 right-handed leptons, 2 transversely
polarized gluons, 2 transversely polarized U(1) gauge bosons, 6
transversely polarized SU(2) bosons, 4 Higgs fields and 2 transversely
polarized states that mix the U(1) and neutral SU(2) boson.  In hard
interactions at such energies, any subsequent
radiation is dominated by emissions that are either soft or
collinear to the colliding or produced particles. 

Processes that only
depend on the flavor of one particle in each of these "jets" of
radiation can be described solely in terms of parton distributions and
fragmentation functions, which have to be evaluated at the
short-distance scale of the hard interaction. The DGLAP evolution of
the PDFs and FFs from the electroweak symmetry breaking scale to the hard
scale $q$ resums the logarithmic dependence on the ratio $m_V / q$. If
the observed particles are not SU(2) singlets, one encounters double
logarithms in the evolution.

We have presented the evolution of FFs in the complete Standard Model,
where all three gauge interactions and the Yukawa interaction of the
third generation contribute significantly to the DGLAP
evolution. Together with the evolution of the PDFs,
which was presented in \cite{Bauer:2016kkv}, this provides all details
necessary to resum the dominant logarithms for all cases where one is
inclusive over the kinematics of the final state particles. 
Combining this with the running of soft functions presented
in~\cite{Manohar:2018kfx}, full NLL accuracy of the electroweak
evolution can be obtained.

While the dominant terms are of double logarithmic origin (scaling as
$\alpha^n L^{2n}$ in a cross section), we also showed how the complete
LL resummation (terms scaling as $\alpha^n L^{n+1}$ in the logarithm
of a cross section) may be achieved by an appropriate choice for the
scale of the running SU(2) coupling in the singular terms of the
evolution. While this does not improve the accuracy in the relevant
limit $\alpha_2 L^2 \sim 1$, and has a small numerical effect on the
resulting FFs, it is necessary when the results from the DGLAP
evolution are combined with the soft function evolution to obtain full
NLL accuracy. 

Numerically, the electroweak logarithms lead to appreciable effects at
the highest energy scales that can be reached at the LHC and a future
100 TeV pp collider, but they still tend to be slightly smaller than
what might be expected from the naive scaling of $\alpha_2 L^2$. For
example, a left handed lepton produced at 3 TeV (30 TeV) has a 6\%
(15\%) probability to fragment into a different particle defined at the
electroweak scale $q_0\sim 100$ GeV. The effect is larger for SU(2)
bosons produced at the high scale.  A charged $W$ boson
produced at 3 TeV (30 TeV) has a 14\% (30\%) probability to fragment
into a different particle defined at 100 GeV.

We have also studied for the first time the phenomenology of
electroweak boson polarization in the FFs.  Although large
polarizations are generated, they have minimal effects as long as the
polarization of fragmentation products is not detected.


\acknowledgments
We thank Aneesh Manohar and Wouter Waalewijn for several discussions about this work.
This work was supported by the Director, Office of Science, Office of
High Energy Physics of the U.S. Department of Energy under the
Contract No. DE-AC02-05CH11231 (CWB), and partially supported by
U.K. STFC consolidated grant ST/P000681/1 (BRW).

\section*{Appendix}
\appendix
\section{Isospin and CP basis}
\label{app:isospin}
As already explained in Section~\ref{sec:intro}, the set of
58 evolution equations can be decoupled to some degree by switching to
a basis of well-defined isospin $\mathbf T$ and ${\mathrm{CP}}$.
Writing a fermion FF with given $\{{\mathbf T},\mathrm{CP}\}$ as
$d_i^{\mathbf T\mathrm{CP}}$, the left-handed fermion FFs are
\begin{align}
\label{eq:fLIsospin}
d^{0\pm}_{f_L} &= \frac
                 14\left[\left(d_{u_L}+d_{d_L}\right)\pm\left(d_{{\bar
                 u}_L}+d_{{\bar d}_L}\right)\right],\\
d^{1\pm}_{f_L} &= \frac
                 14\left[\left(d_{u_L}-d_{d_L}\right)\pm\left(d_{{\bar
                 u}_L}-d_{{\bar d}_L}\right)\right],\\
\,,
\end{align}
where $u_L$ and $d_L$ refer to left-handed up- and down-type
fermions. Right-handed fermion FFs are given by
\beq
\label{eq:fRIsospin}
d^{0\pm}_{f_R} = \frac 12\left(d_{f_R}\pm d_{{\bar f}_R}\right)\,.
\eeq
The SU(3) and U(1) boson FFs have ${\mathbf T} = 0$, with the
unpolarized and helicity asymmetry combinations having ${\mathrm{CP}}=+$
and $-$, respectively:
\begin{align}
d^{0\pm}_g &= d_{g_+}\pm d_{g_-}\,,
     &d^{0\pm}_B &= d_{B_+}\pm d_{B_-}\,.
\end{align}
The SU(2) bosons can have $\{\mathbf T,{\mathrm{CP}}\} = \{0,+\},
\{1,-\}, \{2,+\}$ for the unpolarized FFs and  $\{0,-\}, \{1,+\},
\{2,-\}$ for the asymmetries:
 \beqn
d^{0\pm}_W&=& \frac 13\left[\left(d_{W_+^+}+d_{W_+^-}+d_{W_+^3}\right)
\pm\left(d_{W_-^+}+d_{W_-^-}+d_{W_-^3}\right)\right],\\
d^{1\pm}_W &=& \frac 12\left[\left(d_{W_+^+}-d_{W_+^-}\right)
\mp\left(d_{W_-^+}-d_{W_-^-}\right)\right],\\
d^{2\pm}_W&=& \frac 16\left[\left(d_{W_+^+}+d_{W_+^-}-2d_{W_+^3}\right)
\pm\left(d_{W_-^+}+d_{W_-^-}-2d_{W_-^3}\right)\right].
\eeqn
The mixed $BW$ boson FFs are a combination of $0^-$ and $1^-$ states, and
therefore they have the opposite CP to the corresponding $W$ boson FFs: 
\beq
d^{1\pm}_{BW} = d_{BW_+}\pm d_{BW_-}\,.
 \eeq
For the Higgs boson, one writes similarly to the fermions
 \beqn\label{eq:HIsospin}
&&d^{0\pm}_H = \frac
14\left[\left(d_{H^+}+d_{H^0}\right)\pm\left(d_{H^-}
      +d_{\bar H^0}\right)\right],\\
&&d^{1\pm}_H = \frac
14\left[\left(d_{H^+}-d_{H^0}\right)\pm\left(d_{H^-}
      -d_{\bar H^0}\right)\right].
\eeqn
In terms of these, the longitudinal vector boson and
Higgs FFs are then
\begin{align}
d_{W^+_L} &=d_H^{0+}+d_H^{1+}+d_H^{0-}+d_H^{1-}\,,\\
d_{W^-_L} &=d_H^{0+}+d_H^{1+}-d_H^{0-}-d_H^{1-}\,,\\
d_{Z_L}&=d_h=d_H^{0+}-d_H^{1+}\,.
\end{align}

The resulting evolution equations are collected in Appendix~\ref{app:forward}.

\section{Equations used in the forward evolution}
\label{app:forward}
As in~\cite{Bauer:2017isx},  we define
\beq\label{eq:convol}
P^R_{ji,I}\otimes d^k_j = \int_x^{z_{\rm max}^{ji,I}(q)} \!\!\! \df z \,
P^R_{ji, I}(z) d^k_j(x/z, q)\,.
\eeq
The `+'-prescription is
\beqn
\label{eq:Pplusdef}
P^+_{VV, I} \otimes d_i &\equiv& \left(P^R_{V_+V_+, I}+P^R_{V_+V_-, I}\right)\otimes d_i
+\frac{P^V_{i,I}}{C_{i,I}}d_i\,\\
P^+_{ff, I} \otimes d_i &\equiv& P^R_{ff, I}\otimes d_i
+\frac{P^V_{i,I}}{C_{i,I}}d_i\,\\
P^+_{HH, I} \otimes d_i &\equiv& P^R_{HH, I}\otimes d_i
+\frac{P^V_{i,I}}{C_{i,I}}d_i\,
\eeqn
where $C_{i,I}$ is the coefficient in the corresponding Sudakov factor:
\beqn
\Delta_{i,I}(q)&=&\exp\left[ \int_{q_0}^q \frac{\df q'}{q'}
  \frac{\alpha_I(q')}{\pi} P^V_{i,I}(q') \right]\nn
&=&\exp\left[ -C_{i,I}\int_{q_0}^q \frac{\df q'}{q'}
  \frac{\alpha_I(q')}{\pi} \int_0^{z_{\rm max}^{ii,I}(q)} \!\!\!
  z\,\df z\,P^R_{ii,I}(z)+\ldots\right]\,,
\eeqn
and $\ldots$ represents less divergent terms.  For convenience we also
define the isospin suppression factors
\beq
\Delta^{(T)}_{i}(q) = \left[\Delta_{i,2}(q)\right]^{T(T+1)/C_{i,2}}.
\eeq

For gauge bosons we also need the helicity asymmetry splitting functions:
\beqn
P^A_{VV,I} \otimes d_i&\equiv &\left(P^R_{V_+V_+, I}-P^R_{V_+V_-,
    I}\right)\otimes d_i  +\frac{P^V_{V,I}}{C_{V,I}} d_i\,,\\
P^A_{Vf,I} \otimes d_i&\equiv &\left(P^R_{V_+f, I}-P^R_{V_-f,
    I}\right)\otimes d_i\,,\\
P^A_{fV,I} \otimes d_i&\equiv &\left(P^R_{fV_+, I}-P^R_{fV_-,
    I}\right)\otimes d_i\,,
\eeqn
where
\beqn
P^R_{V_+V_+, G}(z)-P^R_{V_+V_-,G}(z) &=& \frac 2{1-z}+2-4z\,,\\
P^R_{V_+f, G}(z)-P^R_{V_-f,G}(z) &=& z-2\,,\\
P^R_{fV_+, G}(z)-P^R_{fV_-,G}(z) &=& \frac 12-z\,.
\eeqn

\subsection{SU(3) interaction}
\begin{itemize}
\item $\mathbf T = 0$ and ${\mathrm{CP}} = \pm$:
\beqn
\left[ q\frac{\pd}{\pd q} d^{0+}_{q_{L,R}} \right]_3  &=& \frac{\a_3}\pi C_F\left[ P^+_{ff,G} \otimes d^{0+}_{q_{L,R}} 
+P^R_{Vf,G}\otimes d^{0+}_g\right],\\
 \left[ q\frac{\pd}{\pd q}d^{0+}_g \right]_3  &=& \frac{\a_3}\pi\left[C_A P^+_{VV,G}\otimes d^{0+}_g+ T_R
P^R_{fV,G}\otimes d^{0+}_{\sum_g}\right],\\
\left[ q\frac{\pd}{\pd q} d^{0-}_{q_{L,R}} \right]_3  &=& \frac{\a_3}\pi C_F\left[ P^+_{ff,G} \otimes d^{0-}_{q_{L,R}}
\pm P^A_{Vf,G}\otimes d^{0-}_g\right],\\
 \left[ q\frac{\pd}{\pd q}d^{0-}_g \right]_3  &=& \frac{\a_3}\pi\left[C_A P^A_{VV,G}\otimes d^{0-}_g+ T_R
P^A_{fV,G}\otimes d^{0-}_{\sum_g}\right]\,.
\eeqn
Here
\beq
d^{0\pm}_{\sum_g} = 4 \sum_{q_L} d^{0\pm}_{q_L}\pm 2 \sum_{q_R} d^{0\pm}_{q_R}
\,,\eeq
where the sums run over all left-handed quark doublets and all right-handed quarks. The factors of $4$ and $2$ are due to the different normalizations in \eqs{fLIsospin}{fRIsospin}. 

\item All other states:
\beqn
\left[ q\frac{\pd}{\pd q} d_q \right]_3  &=& \frac{\a_3}\pi C_F P^+_{ff,G} \otimes d_q
\,.\eeqn
\end{itemize}

\subsection{U(1) interaction}

\begin{itemize}
\item $\mathbf T = 0$ and ${\mathrm{CP}} = +$:
\beqn
\left[ q\frac{\pd}{\pd q} d^{0+}_f \right]_1 &=& \frac{\a_1}\pi Y_f^2\left[P^+_{ff,G}\otimes d^{0+}_f
+ P^R_{Vf,G}\otimes d^{0+}_B\right],\\
\left[ q\frac{\pd}{\pd q} d^{0+}_B \right]_1 &=& \frac{\a_1}\pi\left[P^V_{B,1} d^{0+}_B +
P^R_{fV,G}\otimes d^{0+}_{\sum_B f} + P^R_{HV,G} \otimes d^{0+}_H \right]\,,\\
\left[ q\frac{\pd}{\pd q} d^{0+}_H \right]_1 &=& \frac{\a_1}\pi \frac{1}{4} \left[
P^+_{HH,G}\otimes d^{0+}_H + P^R_{VH,G} \otimes d^{0+}_B\right]\,,
\eeqn
where
\beq
d^{0\pm}_{\sum_B f} = 4\sum_{f_L} N_f Y_{f_L}^2 d^{0\pm}_{f_L}\pm 2
\sum_{f_R} N_f Y_{f_R}^2d^{0\pm}_{f_R}
\,.
\eeq

\item $\mathbf T = 0$ and ${\mathrm{CP}} = -$:
\beqn
\left[ q\frac{\pd}{\pd q} d^{0-}_{f_{L,R}} \right]_1 &=& \frac{\a_1}\pi Y_f^2\left[P^+_{ff,G}\otimes d^{0-}_{f_{L,R}}
\pm P^A_{Vf,G}\otimes d^{0-}_B\right],\\
\left[ q\frac{\pd}{\pd q} d^{0-}_B \right]_1 &=& \frac{\a_1}\pi\left[P^V_{B,1} d^{0-}_B +
P^A_{fV,G}\otimes d^{0-}_{\sum_B f}\right]\,,\\
\left[ q\frac{\pd}{\pd q} d^{0-}_H \right]_1 &=& \frac{\a_1}\pi \frac{1}{4} 
P^+_{HH,G}\otimes d^{0-}_H \,.
\eeqn

\item ${\mathbf T} = 1$ and ${\mathrm{CP}} = +$:
\beq
\left[ q\frac{\pd}{\pd q} d^{1+}_{BW} \right]_1 =\frac{\a_1}\pi \frac 12
P^V_{B,1} d^{1+}_{BW}
\,.\eeq

\item ${\mathbf T} = 1$ and ${\mathrm{CP}} = -$:
\beq
\left[ q\frac{\pd}{\pd q} d^{1-}_{BW} \right]_1 =\frac{\a_1}\pi \frac 12
P^V_{B,1} d^{1-}_{BW}
\,.\eeq

\item All other states:
\beqn
\left[ q\frac{\pd}{\pd q} d_f \right]_1 &=& \frac{\a_1}\pi Y_f^2P^+_{ff,G}\otimes d_f
,\\
\left[ q\frac{\pd}{\pd q} d_H \right]_1 &=& \frac{\a_1}\pi \frac{1}{4} 
P^+_{HH,G}\otimes d_H \,.
\eeqn

\end{itemize}

\subsection{SU(2) interaction}

\begin{itemize}
\item $\mathbf T = 0$ and ${\mathrm{CP}} = +$:
\beqn\label{eq:SU2f0plus}
\left[ q\frac{\pd}{\pd q} d^{0+}_{f_L} \right]_2 &=& \frac{\a_2}{\pi}\frac 34\left[
  P^+_{ff,G}\otimes d^{0+}_{f_L}+ P^R_{Vf,G}\otimes d^{0+}_W\right] \,,\\
\left[ q\frac{\pd}{\pd q} d^{0+}_W \right]_2 &=& \frac{\a_2}\pi\left[2 P^+_{VV,G}\otimes
  d^{0+}_W+\sum_{f_L} N_f P^R_{fV,G}\otimes d^{0+}_{f_L} + P^R_{HV,G}\otimes d^{0+}_H\right] \,,
  \\
\left[ q\frac{\pd}{\pd q} d^{0+}_H \right]_2 &=& \frac{\a_2}\pi
 \frac{3}{4}  \left[ P^+_{HH,G}\otimes d^{0+}_H + P^R_{VH,G} \otimes d^{0+}_W\right]\,.
\eeqn

\item $\mathbf T = 0$ and ${\mathrm{CP}} = -$:
\beqn\label{eq:SU2f0minus}
\left[ q\frac{\pd}{\pd q} d^{0-}_{f_L} \right]_2 &=& \frac{\a_2}{\pi}\frac 34\left[
  P^+_{ff,G}\otimes d^{0-}_{f_L}+ P^A_{Vf,G}\otimes d^{0-}_W\right] \,,\\
\left[ q\frac{\pd}{\pd q} d^{0-}_W \right]_2 &=& \frac{\a_2}\pi\left[2 P^A_{VV,G}\otimes
  d^{0-}_W+\sum_{f_L} N_f P^A_{fV,G}\otimes d^{0-}_{f_L}\right] \,,
  \\
\left[ q\frac{\pd}{\pd q} d^{0-}_H \right]_2 &=& \frac{\a_2}\pi
 \frac{3}{4} P^+_{HH,G}\otimes d^{0-}_H\,.
\eeqn

\item $\mathbf T = 1$ and ${\mathrm{CP}} = +$:
\beqn
\left[ \Delta_{f}^{(1)}q\frac{\pd}{\pd q} \frac{d^{1+}_{f_L}}{\Delta_{f}^{(1)}} \right]_2 &=& \frac{\a_2}{\pi}\left[-\frac 14
  P^+_{ff,G}\otimes d^{1+}_{f_L}+\frac 12 P^A_{Vf,G}\otimes
 d^{1+}_{W}\right] \\
\left[ \Delta_{V}^{(1)}q\frac{\pd}{\pd
    q}\frac{d^{1+}_{W}}{\Delta_{V}^{(1)}} \right]_2 &=&
\frac{\a_2}\pi\left[P^A_{VV,G}\otimes  d^{1+}_{W}+\sum_{f_L} N_f P^A_{fV,G}\otimes
  d^{1+}_{f_L}\right] \\
  \nn
\left[\Delta_{H}^{(1)}q\frac{\pd}{\pd
    q}\frac{d^{1+}_{H}}{\Delta_{H}^{(1)}}\right]_2 &=&
\frac{\a_2}\pi\left[ -\frac{1}{4} P^+_{HH,G} \otimes d^{1+}_{H}
\right]\\
\left[\Delta_{V}^{(1)}q\frac{\pd}{\pd q}\frac{d^{1+}_{BW}}{\Delta_{V}^{(1)}}\right]_2 &=& 0\,.
\eeqn

\item $\mathbf T = 1$ and ${\mathrm{CP}} = -$:
\beqn
\left[ \Delta_{f}^{(1)}q\frac{\pd}{\pd q} \frac{d^{1-}_{f_L}}{\Delta_{f}^{(1)}} \right]_2 &=& \frac{\a_2}{\pi}\left[-\frac 14
  P^+_{ff,G}\otimes d^{1-}_{f_L}+\frac 12 P^R_{Vf,G}\otimes
 d^{1-}_{W}\right] \\
\left[ \Delta_{V}^{(1)}q\frac{\pd}{\pd
    q}\frac{d^{1-}_{W}}{\Delta_{V}^{(1)}} \right]_2 &=&
\frac{\a_2}\pi\left[P^+_{VV,G}\otimes  d^{1-}_{W}+\sum_{f_L} N_f P^R_{fV,G}\otimes
  d^{1-}_{f_L} + P^R_{HV,G}\otimes d^{1-}_{H}\right] \\
  \nn
\left[\Delta_{H}^{(1)}q\frac{\pd}{\pd q}\frac{d^{1-}_{H}}{\Delta_{H}^{(1)}}\right]_2 &=& \frac{\a_2}\pi\left[ -\frac{1}{4} P^+_{HH,G} \otimes d^{1-}_{H} + \frac{1}{2} \, P^R_{VH,G} \otimes d^{1-}_{W} \right]\,\\
\left[\Delta_{V}^{(1)}q\frac{\pd}{\pd
    q}\frac{d^{1-}_{BW}}{\Delta_{V}^{(1)}}\right]_2 &=& 0\,.
\eeqn

\item $\mathbf T = 2$ and ${\mathrm{CP}} = +$:
\beqn\label{eq:SU2f2plus}
\left[ \Delta_{V}^{(2)}q\frac{\pd}{\pd q} \frac{d^{2+}_{W}}{\Delta_{V}^{(2)}}\right]_2 &=& -\frac{\a_2}\pi P^+_{VV,G}\otimes
  d^{2+}_{W}\,.
\eeqn

\item $\mathbf T = 2$ and ${\mathrm{CP}} = -$:
\beqn\label{eq:SU2f2minus}
\left[ \Delta_{V}^{(2)}q\frac{\pd}{\pd q} \frac{d^{2-}_{W}}{\Delta_{V}^{(2)}}\right]_2 &=& -\frac{\a_2}\pi P^A_{VV,G}\otimes
  d^{2-}_{W}\,.
\eeqn

\end{itemize}

\subsection{Yukawa interaction}
\begin{itemize}
\item $\mathbf T = 0$ and ${\mathrm{CP}} = +$:
\beqn
\left[q\frac{\pd}{\pd q}  d^{0+}_{q^3_L} \right]_Y &=&
\frac{\a_Y}{\pi}\biggl[P^V_{q^3_L,Y} d^{0+}_{q^3_L} + 
P^R_{ff,Y} \otimes d^{0+}_{t_R}  + P^R_{Hf,Y} \otimes d^{0+}_{H} \biggr]\\
\left[q\frac{\pd}{\pd q}  d^{0+}_{t_R} \right]_Y &=&
\frac{\a_Y}{\pi}\,2\,\biggl[P^V_{t_R,Y} d^{0+}_{t_R} + 
P^R_{ff,Y} \otimes d^{0+}_{q^3_L}  + P^R_{Hf,Y} \otimes d^{0+}_{H}\biggr]\\
\left[ q\frac{\pd}{\pd q} d^{0+}_{H} \right]_Y &=&
\frac{\a_Y}{\pi}\biggl[P^V_{H,Y} d^{0+}_{H} + 
N_c P^R_{fH,Y} \otimes d^{0+}_{\sum_H f}\biggr]
\,,\eeqn
where
\beq
d^{0+}_{\sum_H f} = d^{0+}_{t_R} + d^{0+}_{q_L^3} \,.
\eeq

\item $\mathbf T = 0$ and ${\mathrm{CP}} = -$:
\beqn
\left[q\frac{\pd}{\pd q}  d^{0-}_{q^3_L} \right]_Y &=&
\frac{\a_Y}{\pi}\biggl[P^V_{q^3_L,Y} d^{0-}_{q^3_L} + 
P^R_{ff,Y} \otimes d^{0-}_{t_R}  - P^R_{Hf,Y} \otimes d^{0-}_{H} \biggr]\\
\left[q\frac{\pd}{\pd q}  d^{0-}_{t_R} \right]_Y &=&
\frac{\a_Y}{\pi}\,2\,\biggl[P^V_{t_R,Y} d^{0-}_{t_R} + 
P^R_{ff,Y} \otimes d^{0-}_{q^3}  +  P^R_{Hf,Y} \otimes d^{0-}_{H}\biggr]\\
\left[ q\frac{\pd}{\pd q} d^{0-}_{H} \right]_Y &=&
\frac{\a_Y}{\pi}\biggl[P^V_{H,Y} d^{0-}_{H} + 
N_c P^R_{fH,Y} \otimes d^{0-}_{\sum_H f}\biggr]
\,,\eeqn
where
\beq
d^{0-}_{\sum_H f} = d^{0-}_{t_R} - d^{0-}_{q_L^3} \,.
\eeq

\item $\mathbf T = 1$ and ${\mathrm{CP}} = +$:
\beqn
\left[ q\frac{\pd}{\pd q} d^{1+}_{q^3_L} \right]_Y &=&
\frac{\a_Y}{\pi}\biggl[P^V_{q^3_L,Y} d^{1+}_{q^3_L} - P_{Hf,Y} \otimes d^{1+}_{H}\biggr] \\
\left[ q\frac{\pd}{\pd q} d^{1+}_{H} \right]_Y &=&
\frac{\a_Y}{\pi}\biggl[P^V_{H,Y} d^{1+}_{H}  -
N_c P^R_{fH} \otimes d^{1+}_{q_L^3}\biggr]
\eeqn

\item $\mathbf T = 1$ and ${\mathrm{CP}} = -$:
\beqn
\left[ q\frac{\pd}{\pd q} d^{1-}_{t_L} \right]_Y &=&
\frac{\a_Y}{\pi}\biggl[P^V_{t_L,Y} d^{1-}_{t_L} +
P_{Hf,Y} \otimes d^{1-}_{H}\biggr] \\
\left[ q\frac{\pd}{\pd q} d^{1-}_{H} \right]_Y &=&
\frac{\a_Y}{\pi}\biggl[P^V_{H,Y} d^{1-}_{H} + 
N_c P^R_{fH,Y} \otimes d^{1-}_{q_L^3}\biggr]
\eeqn
\end{itemize}

\subsection{Mixed interaction}
\begin{itemize}
\item $\mathbf T = 1$ and ${\mathrm{CP}} = +$:
\beqn
\left[ q\frac{\pd}{\pd q} d^{1+}_{f} \right]_M &=& \frac{\a_M}\pi \frac{Y_f}{2}
P^R_{Vf,G}\otimes d^{1+}_{BW}\,,\\
\label{eq:mixBW1}
\left[q\frac{\pd}{\pd q} d^{1+}_{BW}\right]_M &=& \frac{\a_M}\pi
\left[
4\sum_{f_L} Y_f N_f P^R_{fV,G}\otimes d^{1+}_{f}  + 2 P^R_{HV,G} \otimes d^{1+}_{H}
\right]\,,\\
\left[ q\frac{\pd}{\pd q} d^{1+}_H \right]_M &=& \frac{\a_M}\pi \frac{1}{4} P^R_{VH,G} \otimes d^{1+}_{BW}\,.
\eeqn

\item $\mathbf T = 1$ and ${\mathrm{CP}} = -$:
\beqn
\left[ q\frac{\pd}{\pd q} d^{1-}_{f_L} \right]_M &=& \frac{\a_M}\pi \frac{Y_f}{2}
P^A_{Vf,G}\otimes d^{1-}_{BW}\,,\\
\label{eq:mixBW2}
\left[q\frac{\pd}{\pd q} d^{1-}_{BW}\right]_M &=& \frac{\a_M}\pi
4\sum_{f_L} Y_f N_f P^A_{fV,G}\otimes d^{1-}_{f} \,,\\
\left[ q\frac{\pd}{\pd q} d^{1-}_H \right]_M &=& 0\,.
\eeqn

\end{itemize}
Equation~(\ref{eq:mixBW1}) differs slightly from
Ref.~\cite{Ciafaloni:2005fm} where, taking into account the definition
there of $d_{B3}=d_{BW}/2$, an 8 would appear in place of 4 in the first term on
the right-hand side.

\addcontentsline{toc}{section}{References}
\bibliographystyle{JHEP}
\bibliography{SMevol_paper}

\end{document}